\documentclass[journal]{IEEEtran}
\usepackage{amsmath,amsthm,amssymb,amsfonts}
\usepackage{tikz}
\newcommand{\pgftextcircled}[1]{
    \setbox0=\hbox{#1}%
    \dimen0\wd0%
    \divide\dimen0 by 2%
    \begin{tikzpicture}[baseline=(a.base)]%
        \useasboundingbox (-\the\dimen0,0pt) rectangle (\the\dimen0,1pt);
        \node[circle,draw,outer sep=0pt,inner sep=0.1ex] (a) {#1};
    \end{tikzpicture}
}
\newtheorem{theorem}{Theorem}
\newtheorem{definition}[theorem]{Definition}
\newtheorem{proposition}[theorem]{Assumption}
\usepackage{amsmath,amsthm,amssymb,amsfonts}
\usepackage{graphicx}
\usepackage{amsthm}
\usepackage{amsmath}
\usepackage{booktabs}
\usepackage{multirow}
\usepackage{algorithm}
\usepackage{xcolor}
\usepackage[utf8x]{inputenc}
\usepackage[T1]{fontenc}
\usepackage{subfigure}
\usepackage{algorithmic}%[noend]
\usepackage{amsfonts}
\usepackage{mathrsfs}
\usepackage{amssymb}
\usepackage{threeparttable}
\usepackage{graphicx}
\usepackage{url}
\usepackage{multicol}
\begin{document}
\renewcommand{\algorithmiccomment}[1]{//#1}
\renewcommand{\algorithmicrequire}{\textbf{Input:}}
\renewcommand{\algorithmicensure}{\textbf{Output:}}
\title{CrowdExpress: A Probabilistic Framework for On-Time Crowdsourced Package Deliveries}

\author{Chao~Chen, Sen Yang, Weichen Liu, Yasha Wang, Bin Guo and Daqing Zhang
        
\IEEEcompsocitemizethanks{
\IEEEcompsocthanksitem Chao Chen, Sen Yang and Weichen Liu are with the Key Laboratory of Dependable Service Computing in Cyber Physical Society (Chongqing University), Ministry of Education and also with the College of Computer Science, Chongqing University, Chongqing 400044, China. E-mail: cschaochen@cqu.edu.cn.
\IEEEcompsocthanksitem   Yasha Wang and Daqing Zhang are with the Institute of Software, School of Electronics Engineering and Computer Science, Peking University, Beijing 100871, China. E-mail: \{dqzhang,wangys\}@sei.pku.edu.cn.
\IEEEcompsocthanksitem Bin Guo is with the School of Computer Science,  Northwestern Polytechnical University, Xi'an 710072, China. E-mail: guobin.keio@gmail.com.
%\IEEEcompsocthanksitem Xu Wang is with College of Mechanical Engineering, Chongqing University, Chongqing 400044, China. E-mail: wx921@163.com.
}
}

%\markboth{IEEE TRANSACTION ON INTELLIGENT TRANSPORTATION SYSTEMS, 2012}%
%{Shell \MakeLowercase{\textit{ Chen et al.}}: iBOAT}

%\IEEEcompsoctitleabstractindextext{%
\maketitle

\begin{abstract}
Speed and cost of logistics are two major concerns to on-line shoppers, but they generally conflict with each other in nature. To alleviate the contradiction, we propose to exploit  existing taxis that are transporting passengers on the street to relay packages collaboratively, which can simultaneously lower the cost and accelerate the speed. Specifically, we propose a probabilistic framework containing two phases called CrowdExpress for the on-time package express deliveries. In the first phase, we mine the historical taxi GPS trajectory data {\em offline} to build the package transport network. In the second phase, we develop an {\em online} adaptive taxi scheduling algorithm to find the path with the maximum arriving-on-time probability ``on-the-fly'' upon real-time requests, and direct the package routing accordingly.  Finally, we evaluate the system  using the real-world taxi data generated by over 19,000 taxis in a month in the city of New York, US. Results show that  around 9,500 packages can be delivered successfully on time per day with the success rate over 94\%, moreover, the average computation time is within 25 milliseconds.  
\end{abstract}
%}
% make the title area
\IEEEpeerreviewmaketitle

\section{Introduction}
\IEEEPARstart{T}he pervasive use of tablet and mobile devices leads to increased popularity of round-the-clock online shopping, which urges the sustainable development of logistics industry~\cite{lastmile}.  For instance, online ordered products generated over one billion package deliveries in 2013, and this number is predicted to grow by 28.8\% in 2018~\cite{lastmile}. To on-line shoppers, {\em speed and cost} are two major concerns, between which they pay relatively more attention to cost~\cite{tan14express}.  Unfortunately, speed and cost  usually  conflict with each other in nature. In another word, speeding up the shipping process often implies more personnel and vehicles on the road, incurring more extra cost as a result~\cite{arslan2018crowdsourced}.  For example, users have to pay a high amount of money to enjoy the express service, such as 5  dollars per package if users want the {\em same-day delivery} service in US~\cite{economist14sameday}. Therefore, logistics services which can lower the cost but still ensure the arrivals of packages on time are preferable~\cite{Rohm04shoppers}.

% Online retailing research has shown that most people shop online for convenience~\cite{Rohm04shoppers}. Thus,  more and more online retailers and logistics providers  offer the {\em same-day delivery} service  to attract shoppers~\cite{economist14sameday}. For example, the products can be delivered no later than 5:00 pm if consumers made online orders before 9:00 am (i.e., ensuring the products can arrive at their destinations on time). 
With the sustainable development and proliferated daily use of positioning and mobile Internet technologies, rich data regarding  status of the vehicles, passengers, and packages (e.g., real-time positions, origins and destinations of passengers and packages, available transport capacity of a vehicle) can be easily recorded and accessible in real time~\cite{chen2016promises,zhang2011emergence}. In this context, crowdshipping (also termed as crowd logistics, crowdsourcing logistics) which receives the increasing attention from both academic and industrial communities in the last few years, has been recognized as a {\em promising and cost-effective} way to alleviate the contradiction by sending  passengers and packages simultaneously in a shared space and transport network~\cite{chen16crowddeliver,devari16crowdsourced,fagnant2014travel,li2014share,rai2017crowd}. In line with the previous research, with a particular focus on taxis,  we propose having packages take {\em hitchhiking rides} collaboratively with existing taxis that are transporting passengers on the street, i.e., the existing mobility of taxi drivers~\cite{arslan2018crowdsourced,liu2018foodnet}. We illustrate the basic idea using the following  example.% as shown in Fig.~\ref{fig:example1}.

{\em There is a package to be delivered from \textbf{A} to \textbf{B}. Here we simply assume that both \textbf{A} and \textbf{B} have facilities such as smart parcel boxes that can store packages temporally. Opportunistically, a passenger (e.g. $passenger_1$) at \textbf{A}  makes a real-time taxi ordering request\footnote{It is popular that passengers order taxis in real time with mobile apps such as Uber (\url{https://www.uber.com/}). Usually, to make a request, a passenger has to provide information including his/her origin, intended destination. The request will be broadcast locally, and the taxi driver who  accepts request would come to pick up the passenger~\cite{leng15taxiapps}.},  intending to go to  \textbf{B}. Once a taxi (e.g. $taxi_1$) responds to the request, we can assign the package delivery task to its driver.  More specifically, we can  ask the taxi driver to first collect the package and then pick up the passenger. After dropping off  $passenger_1$ at \textbf{B}, the taxi driver leaves the package at an appointed location. Finally, the package will be delivered to or collected by its receiver at \textbf{B}. }

%\begin{figure}
%\centering
%\includegraphics[width=0.94\columnwidth]{./figs/example.pdf}
%\caption{An example of a package delivery which takes a single hitchhiking ride.}
%\label{fig:example1}
%\end{figure}

As shown in the above example, this approach only requires small additional efforts and time  from the taxi drivers involved, without degrading the service quality and any interrupt to passengers. We can formulate the proposed taxi-based package delivery as a route planning problem  with the objective of delivering the packages to their destination on time (by the deadline given by users).  To make  the  idea of organizing the passenger flow and package flow seamlessly via the taxi transport network feasible, we need to address the following two major research challenges:
\begin{itemize}
    \item Package flow and passenger flow are {\em  incompatible} in time and space. More specifically, 1) compared to package flow,  passenger flow presents salient peak-hour patterns; 2) due to the financial considerations, most passengers choose to  take taxis only when the destination is close, e.g., within 4 km~\cite{chen2014b}. While for packages, the destination is generally  far away from the origin (e.g., longer than 5 km)~\cite{arslan2018crowdsourced}.  Therefore, we argue that   routing algorithms based on the framework of  {\em query-matching} merely may not work~\cite{devari16crowdsourced,li2014share}, since a single hitchhiking ride may not be able to deliver a package to its destination; instead, a collaborative relay of taxis is needed. %even for the last-mile deliveries, partly because the logistics distribution centers are commonly scattered around the outskirt of the city
     \item Requests for demands of packages and passengers come with high {\em uncertainties} in stream. Although regular spatial and temporal patterns about passenger flow have been unveiled from taxi GPS trajectory data in the coarse granularity,  it is challenging to predict the passenger demands accurately in a quite fine granularity (e.g., the passenger demands in the next 5 minutes for a given road segment)~\cite{moreira2013predicting,yao2018deep,zhang2018predicting}. Meanwhile,  same situation also happens to package demands~\cite{tan14express}. In summary, uncertainties exist in both requests, which hinder the estimation and comparison of time cost of package routing paths.
\end{itemize}
With the above-mentioned research objective and challenges, the {\bf main contributions} of the paper are:

\begin{enumerate}
    \item We propose a novel passenger and package mixed transport mode which leverages the {\em unintentional cooperations} among a crowd of occupied taxis to deliver  city-wide packages on time, in order to  lower the transport cost and enhance the transport efficiency simultaneously. %Moreover, package uploading (offloading) takes place by taxi drivers before picking up (after dropping off) passengers, thus the service quality to passengers is not degraded.  i.e., finding the optimal one that is expected have a maximum probability of arriving-on-time ``on-the-fly''  for each package delivery (i.e., the determination of the next stops that the package would head and the taxi scheduling), rather than computing and  comparing the total time cost of each potential delivery route directly. 
    \item We formulate the package routing problem as the arriving-on-time problem~\cite{nie06arriving} to tackle with uncertainties of passenger and package requests. Moreover, we propose a probabilistic  framework named CrowdExpress, which contains two phases to solve it.  In the first phase, we build the package transport network by mining the historical taxi GPS trajectory data {\em offline}. In the second phase,  for each real-time generated package delivery request, we propose an {\em online adaptive taxi scheduling algorithm} based on the probabilistic model called {\bf maxProb} to iteratively determine the next stop of the package ``on-the-fly". The algorithm monitors real-time taxi ordering requests, {\em recursively} computes the  maximum arriving-on-time probability if assigning the delivery task to the currently available taxi, and compares it to the one if waiting for future taxi rides, based on the real-time package location and the remaining time budget.   
    \item We conduct extensive evaluations using  road network data and taxi GPS trajectory data generated over 19,000 taxis in a month in the city of New York (NYC) to verify the efficiency and effectiveness of CrowdExpress. Results demonstrate that CrowdExpress responds within 25 milliseconds. What is more, it can throughput around 9,500 packages daily with the success rate over 94\%, i.e., over 94\% of packages can be delivered successfully on time, which is consistently better than the baseline approaches. 
\end{enumerate}

The rest of the paper is organized as follows. In Section~\ref{sec:relatedwork}, we review the related work and show how this paper differs from prior research. In  Section~\ref{sec:preliminary}, we introduce some basic concepts, the assumptions we have made,  the formal problem formulation, and  overview the system. We present the technical details about our two-phase approach in  Section~\ref{sec:phase1} and Section~\ref{sec:phase2} respectively. We evaluate the performance of the proposed framework in Section~\ref{sec:evaluation}.  Finally, we conclude the paper and chart the future directions in Section~\ref{sec:conclusion}. 
\section{Related Work}\label{sec:relatedwork}
Here, we will review the related work, which can be categorized into two groups. The first group consists of the work on crowdsourced logistics, whereas the second group focuses on taxi trajectory mining and its supporting applications.
\subsection{Crowdsourced Logistics}
Crowdsourcing has been used for many different applications, from problem solving~\cite{Doan11crowdsourcing} to various  sensing tasks~\cite{guo15mobile,semanjski2016crowdsourcing,zhang14wh}. There exist two concrete papers that particularly targeted at the package delivery problem, leveraging the spatial and time overlaps between crowdsourcing workers. Specifically, Sadilek et al.~\cite{sadilek14crowdphysics}  recruited a group of twitter users, asking one person to pass  the assigned package to another twitter user that happened to be nearby (within a certain distance). However, this work  had two main limitations: 1) It is hard to trace and coordinate the users since people rarely share their location information continuously via geo-twitter~\cite{sadilek12geo}. Therefore, the choice of suitable deliverers is limited, probably resulting in longer and uncontrollable package delivery delay. 2)  It may be not practical to ask a participant to make a dedicated trip to pass the package to another suitable user, as it may interrupt his/her on-going activities (e.g. having conversation/dinner with friends) that are hard to be inferred from the user's geo-twitters data. Similarly, the work in~\cite{McInerney14Crowd} which employed mobile users  based on the overlaps of space and time  inferred from cell towers had similar limitations: 1) The cell towers may be sparse in certain areas and thus it is difficult for mobile users to relay packages in that areas. 2) Only with data when people make calls, as a result, the number of mobile users can be recruited is limited. Compared to the proposed solutions in~\cite{McInerney14Crowd,sadilek14crowdphysics}, there are also some papers which intended to leverage the abundant existing passenger-delivery trips to hitchhike packages  appeared in these two years~\cite{arslan2018crowdsourced,archetti2016vehicle,devari16crowdsourced,fatnassi2015planning,kafle2017design,li2014share,li2016share}. Although the passenger flow and package flow are combined to be transported mixed, the authors fail to consider their distinct patterns in time and space. Specifically,  they formulate the problem as the {\em share-a-ride problem} and insert the package requests into the passenger-delivery trips, which may not able to deliver packages successfully in real cases as we argued previously. To make the matters worse, in their solutions, during the passenger-sending course, taxis have to make several dedicated stops and detour, which degrades the service quality to passengers.  {\color{black} At the current stage, the research on crowdsourced logistics mainly focused on issues regarding how to efficiently discover the `optimal' package delivery paths and almost completely ignored the multi-criteria  design of the package relay network~\cite{arslan2018crowdsourced,chen16crowddeliver,chen2017multi,kafle2017design}. How to design the package relay network (e.g., the optimal number and location of package interchange stations) is vital and challenging, which should be a separated research issue~\cite{ali2002relay,uster2011strategic}. In this paper, we exploit existing taxi services  to deliver packages~\cite{chen2017multi} and simply cluster frequent pickup and drop-off points to get package interchange stations without any optimization technicals.}  Packages can be temporary stored at interchange stations in-between rides, and thus no time overlap and pair-wise contact between participants is needed. Our method requires less effort from the participants and can transport packages over a longer distance.  In addition, we try to minimize the impact on the quality and experience of passenger service, which is similar to our previous work~\cite{chen16crowddeliver}. Generally speaking, CrowdExpress is different from~\cite{chen16crowddeliver} in the {\em objective, proposed solution as well as the evaluation environment}. To be more specific, CrowdExpress aims to send each package to the destination on time (by the deadline required by the user), rather than as quick as possible. We argue that the objective is more reasonable and matches the real-life application scenarios more. In CrowdExpress, we formulate the problem as the one of finding arriving-on-time paths, and propose a probabilistic framework to address it. We evaluate the performance in NYC, using the real-world taxi trajectory data. {\color{black} Due to the openness, we choose to use the taxi trajectory data from NYC in our experiments, expecting to lower the data barrier and attract more researchers with different backgrounds into the promising interdisciplinary filed.}  
%The packages are delivered without any dedicated trips made by taxis, since they are delivered by taking a number of hitchhiking rides provided by  occupied taxis while they are sending passengers.   %

%In this work, we leverage taxis on the street which can travel longer distance. Further more, the packages are stored at the safe-boxes installed at the roads temporally  between rides, and thus no time overlap is needed, requiring less participants' efforts, 

\subsection{Taxi Trajectory Data Mining}
Information mined from taxi trajectory data can benefit for taxi drivers, passengers and city planners. For taxi drivers  who are mostly interested  in making more money while minimizing the fuel cost, many papers have tried to recommend areas with more potential passengers, e.g.~\cite{Ge10kdd,Yuan11T-finder}. In addition, Zhang et al.~\cite{zhang14taxi} investigated  the  differences between {\em efficient and inefficient} taxi service in terms of  passenger-searching, passenger-delivery and service-region preference. Work on recommending the best corner to catch taxis, real-time ordering free taxis, and the taxi fee estimation aims to improve the experiences of passengers, e.g.~\cite{Yuan11T-finder}.  An interesting work detected anomalous taxi rides and warned the passengers ``on-the-fly'' that they were taken on a unnecessary detour~\cite{Chen13iBOAT}.  For city planners, taxi trajectory data provides a rich data source to identify flaws in city planning~\cite{Liu14bus,zheng11urban}, probe traffic conditions~\cite{castro12urbantraffic}, estimate the travel demands, infer the land-use efficiency~\cite{Liu12urban,Pan12Landuse}, suggest bus routes~\cite{Liu14bus}, etc. Castro et al.~\cite{Castro14Survey} provided a good survey on understanding city dynamics  using taxi trajectory data. Recent studies also incorporate taxi trajectory data  with other data sources such as POI data, Foursquare check-in data, and Flickr image data, to enable smarter applications, such as  building functions inferring, personalized and fuel-efficient travel route planning, real-time travel purpose inference and so on~\cite{chen2018tripimputor,Chen15trip, ding2017greenplanner,pan13crowd,yu16personalized,Zhong14function}. {\color{black} Another stream is to revisit previous well-known research issues using the advanced machine learning skills. For instance, Xu et al.~\cite{xu2017real} forecasted taxi demands in real time using recurrent neural networks. Tong et al.~\cite{tong2017simpler} applied a simple linear regression model with more than 200 million dimensions of features  to predict the original taxi demands for each POI.} However, most of the  existing work that leverages taxi trajectory data focuses on transporting passengers. Little attention has been paid to shipping goods. To the best of our knowledge, we are among the first to target this new application.

\section{Preliminary, Problem Statement and System Overview}\label{sec:preliminary}
%{Basic Concepts, Assumptions, Problem Statement and System Overview}
In this section, we provide definitions of some basic concepts, elicit assumptions we have made, and give a formal problem statement.  Finally, we give an overview of the proposed CrowdExpress system. 
\subsection{Preliminary}
\subsubsection{Basic Concepts}
We define the basic concepts used in this work as follows:

\begin{definition}[Taxi Trajectory]
A taxi trajectory $tr$ is a sequence of GPS points corresponding to a single passenger-delivery trip. Here, the taxi trajectory is represented by a pair of Origin-Destination (OD), where the origin is the road segment that the trip starts and the destination is the one that the trip ends.  The travel time is exactly the time difference between the ending and starting times. %Due to privacy issues, the detailed travel routes between OD pairs are not provided. 
\end{definition}

\begin{definition} [Real-time Taxi Ordering Request]
A taxi ordering request  ($tor$) is defined as a triplet  $\langle o_t,d_t,r_t\rangle$, where $o_t$ and $d_t$ refer to the passenger's origin and intended destination, respectively.  $r_t$ refers to the time that the passenger submits the request.
\end{definition}

\begin{definition}[Package Transport Network] 
A package transport network is a graph $G(S,E)$, consisting of a node set $S$ and an edge set $E$, where each element $s$ in $S$ is an interchange station which is responsible for package collections and storage.  Edge set $E$ is a subset of the cross product $S\times S$. Each element $e(i,j)$ in $E$ is a non-stop directional transport route from node $s_i$ to node $s_j$, implying that there is an abundant passenger flow for hitchhiking packages. {\color{black} It should be noted that the edge in the package transport network has different meaning from the edge defined over the road network. There can be multiple driving paths over the road network connecting two interchange stations.} % which is associated with the travel time. 
\end{definition}

\begin{definition}[Package Delivery Request]
A package delivery request ($pr$) is defined as a triplet $\langle o_p,d_p,t_p \rangle$, where $o_p$ and $d_p$ refer to the origin and destination of the package delivery respectively; $t_p$ refers to the time when the user submits the request (i.e. the birth time). Note that here $o_p\in S$, $d_p\in S$, indicating that packages should originate and end at interchange stations. 
\end{definition}

\begin{figure}
	\centering
	\includegraphics[width=0.984\columnwidth]{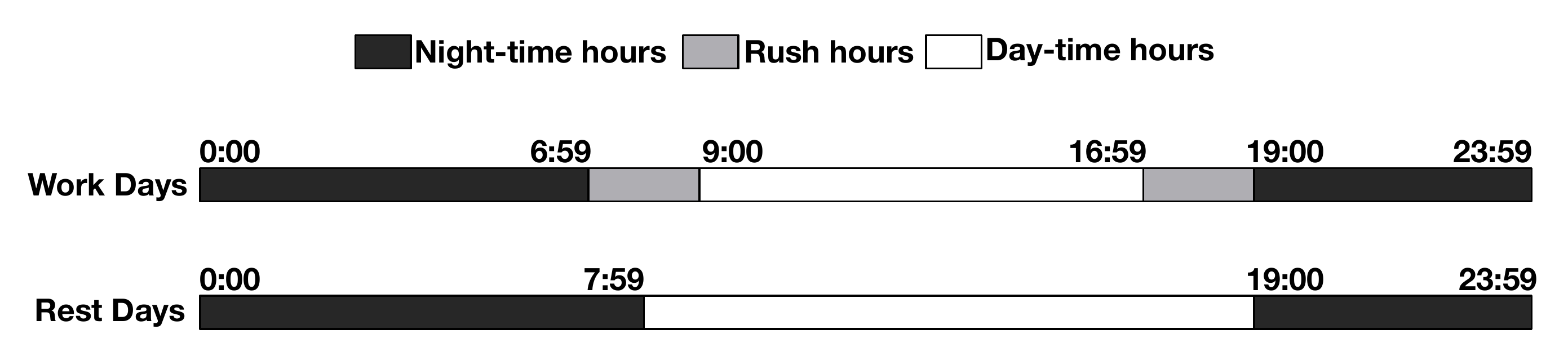}
	\caption{Time slot slicing.}
	\label{fig:time}
\end{figure}

\begin{figure}
    \centering
    \includegraphics[width=0.884\columnwidth]{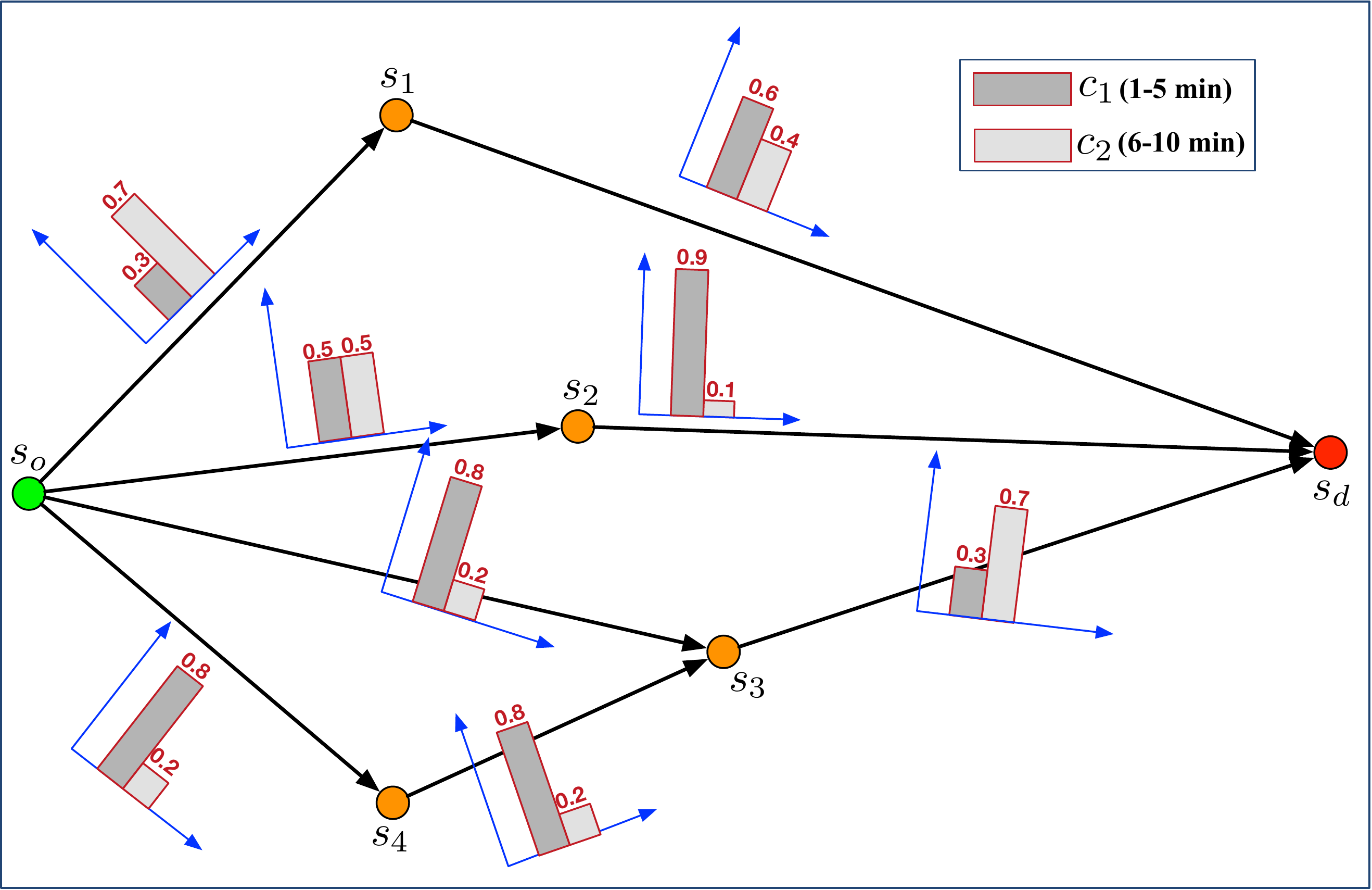}
    \caption{An illustrative example of package delivery paths from $s_o$ to $s_d$, as well as the distribution of discrete travel times on each edge.}
    \label{fig:example}
\end{figure}

{\color{black}\begin{definition}[Time Slot Slicing]
We divide a whole day into different time slots (periods) according to the day type, since the traffic conditions are changing in different time slots, resulting in large variance in travel time. A work day is divided into three time slots and a rest day is divided into two time slots, as detailed in Fig.~\ref{fig:time}.
\end{definition}
}
{\color{black}
\begin{definition}[Travel Time Probability Function] 
Each edge in the package transport network ($G$) is associated with an independent random travel time (cost) $t_{ij}$ whose probability density function is denoted by $p_{ij}(t)$.  $p_{ij}(t)$ varies at different time slots.   %We category all travel times of passenger-sending trips into several time-bins manually, as shown in Table~\ref{tab:timebins}. 
\label{def:probability}
\end{definition}
For instance, the probability that a package spends a time in the interval [$0,t_0$] from node $s_i$ to node $s_j$ directly can be computed by the definition, as shown in Eq.~\ref{eq:integral}. 
\begin{equation}
    P_{ij}(t\le t_0) = \int_{0}^{t_0} p_{ij}(t)dt
    \label{eq:integral}
\end{equation}
The travel time probability function in each time slot can be obtained separately, according to Eq.~\ref{eq:integral}. What is more, as can be observed, the travel time probability is a {\em monotonic increasing function} of the time $t$.
}
\begin{definition}[Travel Time Discretization] To simplify the calculation of travel time probability along an edge, we consider the travel time in a discrete manner. More precisely, we use a piecewise constant function with equal step width $\tau$  to  discretize  different travel times\footnote{Here, we set $\tau=5~min$ throughout the whole paper.}. % Moreover, travel times greater than 40 minutes occupy quite a small percentage, i.e., less than 1.38\% in the taxi trajectory data in the city of New York (NYC), thus they are excluded in advance.
\label{def:dis} 
\end{definition}
In the discrete case,  the  integral shown in Eq.~\ref{eq:integral} can be replaced using the formula shown as follows.
\begin{equation}
    P_{ij}(t\le t_0)= \frac{\# travel times<\alpha\tau}{\# travel times} \nonumber
\end{equation} 
\begin{equation}
    t_0 = (\alpha-1) \tau +\delta   
\end{equation}
where $\# travel times$ refers to the number of all the possible travel times from node $i$ to node $j$ which are recorded in the taxi trajectory in history, while $\# travel times < \alpha \tau$ refers to the number of travel times less than $\alpha \tau$ after time discretization as defined; $\alpha\in N^+$, and $0\le\delta<\tau$. For the edge from $s_o$ to $s_1$ shown in Fig.~\ref{fig:example}, $P_{o1}(t\le 5) = 0.3; P_{o1}(5<t\le 10)=0.7$.
%Note that travel time of the same OD pair can consist of a number of travel time-bins since the it varies due to different time and traffic condition. Thus, for each edge in the package transport network ($G$), we actually can use the distribution of travel time-bins to represent it, as illustrated in Fig.~\ref{fig:example}. 
%
%\begin{table}[h]
%\centering
%\caption{Travel time-bins of taxi trips}
%\label{tab:timebins}
%\begin{tabular}{c|c||c|c||c|c}
%\toprule
%Category & Travel time     & Category & Travel time & Category & Travel time         \\
%\hline
%$c_1$       & 1-5 min         & $c_4$    & 16-20 min   & $c_7$       & 31-35 min           \\
%$c_2$       & 6-10 min        & $c_5$       & 21-25 min   & $c_8$       & 36-40 min           \\
%$c_3$       & 11-15 min       & $c_6$       & 26-30 min   & $c_9$       & \textgreater 40 min\\
%\bottomrule
%\end{tabular}
%\end{table}

\begin{definition}[Arriving-on-Time Probability] The arriving on time probability of a package-delivering path within a given time duration (i.e., deadline, $t_0$)  is defined as the ratio of the number of travel times less than $t_0$ to the number of all possible travel times (suppose that the package is shipped from $s_i$ to $s_j$ via $s_k$), as follows. 
\begin{equation}
    P_{ij}(t\le t_0|Path) = \int\limits_{t_0-t_1}^{t_0}dt_2\int\limits_{0}^{t_1} p_{ik}(t_1)p_{kj}(t_2-t_1)dt_1
    \label{eq:path}
\end{equation}
\end{definition}

It is obviously that the integral computation becomes more complicated if the given path is longer (i.e., contains more interchange stations) as more travel time combinations are generated.

Similarly, to ease the computation of integral in Eq.~\ref{eq:path}, we first let the travel time be considered in the discrete manner, then the integral can be degraded to the sum computation. Taking the path ($path_1:s_o\rightarrow s_1\rightarrow s_d$) shown in Fig.~\ref{fig:example} as an example, its arriving-on-time probability within 15 minutes can be computed as: 
\begin{equation}
P_{ij}(t\le15|path_1)= \underbrace{0.3\times (0.6+0.4)}_{\text{Case I}} + \underbrace{0.7\times 0.6}_{\text{Case II}} = 0.72\nonumber   \label{eq:pij}
\end{equation}
For the given path, two cases can lead to the successful arrival of packages by the deadline. {\bf Case I:}  If the first recruited taxi driver spent no more than 5 minutes in the first segment (i.e., $s_o\rightarrow s_1$), due to the sufficient time margin left,  then the second recruited taxi driver can arrive at $s_d$ by the deadline {\em  at a hundred percent}. {\bf Case II:} If the first recruited taxi driver took more than 5 minutes in the first segment, then the package can be arrived on time {\em only if} the second recruited taxi driver spent no more than 5 minutes to accomplish the second segment (i.e., $s_1\rightarrow s_d$).
\setcounter{theorem}{0}
\begin{theorem}
For a unique path, its arriving-on-time probability becomes higher (or at least unchanged) if given a longer deadline, mathematically, we have:
\begin{equation}
    P_{ij}(t\le t_1|Path)\le P_{ij}(t\le t_2|Path),~if~t_1<t_2
    \label{eq:theorem}
\end{equation}
\label{th:biggerdeadline}   
\end{theorem}
\vspace{-0.4cm}
\begin{proof} The proof  can be found in Appendix~\ref{sec:appProof}. %easily completed by induction. 
\end{proof}

\setcounter{theorem}{8}
\begin{definition}[Maximum Probability of Arriving-on-Time] For an OD pair, the maximum probability of arriving-on-time (with time cost no greater than $t_0$) is defined as the maximal one among all probabilities on all possible $N$ paths from the origin ($s_i$) to the destination ($s_j$), denoted by $u_{ij}(t\le t_0)$. In another word, $u_{ij}(t\le t_0)$ serves as the upper bound of $P_{ij}(t\le t_0)$.
\end{definition}

If a package at node $s_i$ is firstly sent to $s_k$ next, the probability that the package spends a time in the interval [$\omega,\omega+d\omega$] on edge $s_i,s_k$ is $p_{ik}(\omega)d\omega$, thus the time margin at node $s_k$ is $t_0-w$.  On the basis of {\em Bellman's principle of optimality}~\cite{bellman15applied,nie06arriving}, no matter which node $s_j$ that the package is elected to send next, the package must follow the optimal routing strategy in shipping from node $s_k$ to the destination $s_j$ within the remaining time $t_0-\omega$. Therefore, the maximum probability of arriving-on-time can be formally defined {\em recursively} as follows:  
\begin{equation}
u_{ij}(t\le t_0)=\max\limits_{k\neq j}\int\limits_{0}^{t_0} p_{ik}(\omega)u_{kj}(t\le t_0-\omega)d\omega    
\label{def:maximum}
\end{equation}
%where $s_k$ is the neighouring interchange stations connected to $s_0$.  

Intuitively, according to {\em Definition}~\ref{def:maximum}, to compute the maximum probability of arriving-on-time for an OD pair, {\em one needs to find all possible paths from the origin to the destination}, which is  well-known as an NP-hard problem~\cite{golden1987orienteering}. To make the concept clear, we use the example shown in Fig.~\ref{fig:example} again. It is easy to find all paths from $s_o$ to $s_d$, that is, $path_1:s_o\rightarrow s_1\rightarrow s_d$,  $path_2:s_o\rightarrow s_2\rightarrow s_d$,  $path_3:s_o\rightarrow s_3\rightarrow s_d$ and  $path_4:s_o\rightarrow s_4\rightarrow s_3\rightarrow s_d$, respectively. For each path, similar to the computation in Eq.~\ref{eq:path}, it is not difficult to obtain the respected arriving-on-time probability within a given deadline (e.g., $t_0=15$ minutes). As a result, $u_{ij}(t\le15)= 0.95$ for the example when delivering the package via $path_2$. It is obvious that $u_{ii}(t)=1$ given any deadline, since no travel time is needed if the package stays still. We exclude the round trip in the study. %$u_{ij}(t=\)=1$ if allowing a deadline. 

\subsubsection{Assumptions}
In this work, we make the following assumptions.
\setcounter{theorem}{0}

\begin{proposition}
Taxis cannot be recruited to take part in the package delivery tasks  in the course of sending passengers. We further assume the recruited taxis have enough room for the package storage. 
\end{proposition}

Compared to~\cite{arslan2018crowdsourced,li2014share,li2016share},  the taxis are recruited to collect packages before  picking up passengers, and offload packages after dropping off passengers to minimize potential impact on passengers' experiences. This assumption tries to guarantee the quality of taxi services for passengers.   {\color{black} In fact, it is infeasible that taxi drivers delicately stopped to get packages on-the-way when delivering passengers, unlike to the ride-sharing among passengers. In  the case of passenger ride-sharing,  passengers can be easily picked up on-the-way because they can {\em proactively} wait and get on cars at the appointed locations. On the contrary, in order to get packages, taxi drivers must have to take a  sequence of complex actions, including parking, getting off the car, loading the package, getting on the car, and so on.} %resulting in the infeasibility.  
\begin{proposition}
The taxi drivers are willing to accept the assigned package delivery tasks. 
\end{proposition}

We believe that this assumption can be  realistic given proper incentive mechanisms. In the design of incentive mechanisms,  a prime principle is to ensure  that the reward matches the amount of efforts put in by the drivers. For example, some places are harder to get to or park the taxi, then the incentive should be higher. However, designing a proper incentive mechanism is beyond the scope of this paper. 

\begin{proposition}
The packages are traceable. 
\end{proposition}

In the delivery process, a package is either stored at the interchange station or carried by a scheduled taxi. Each interchange station is authorized and has a unique Id; each taxi is registered in taxi management department and also has a unique Id.  This assumption tries to address the package security issues. Any package damage or loss will impair this novel package delivery service.  

\subsection{Problem Statement}
The collaborative crowdsourced package deliveries leveraging the relays of passenger-occupied taxis can be viewed as the problem of finding arriving-on-time paths, and thus can be  formulated as follows:

{\bf Given:}
\begin{enumerate}
%\item A road network $G(N,E)$ and a set of package interchange stations $CS=\{cs_1,cs_2,\cdots,cs_m\}$ in the designated city,
\item A historical set of taxi trajectory  records $\{Tr\}$, such as from the past month in the designated city,
\item A set of real-time taxi ordering requests $TOR$ from  mobile phone apps, and a set of real-time package delivery requests $PR$. Note that these two requests come in stream,
\item A given deadline specified by the user for each package delivery request.
\end{enumerate}

{\bf Objectives:} {\bf Build} a package transport network (i.e., the identification of interchange stations and estimation of edge values) based on the historical taxi trajectory data. Moreover,  {\bf find} a  package delivery path for each package request ($pr$), which can make the package arrive the destination by the deadline.  However, such package delivery path may not be unique or existed. To migrate the issue, {\em we thus transform the problem to the arriving-on-time problem, i.e., finding the optimal one that is expected to have the maximum probability of arriving-on-time\footnote{If the optimal one is still unable to send the package on-time, then it is safe to claim that the package delivery is an unsuccessful one.}.}

The following two {\bf constraints} should be satisfied.
\begin{enumerate}
\item Only taxis that accept the real-time taxi ordering requests after the package delivery request is posted can be scheduled, i.e. $\{tor.r_t\}>pr.t_p$.
\item A recruited taxi can be available  to participate again only after completing the current task (i.e., dropping off the package at the predefined interchange station). %In other words, a taxi can carry at most one package when sending passengers. 
%\item The number of transshipments is no more than $k$, where $k$ is a user-specified parameter.  
%\item Packages cannot be delivered during rush hours, since the demand of passenger taking taxis is high during that time period, and taxis should ensure its main service to meet passengers' demands primarily.  
\end{enumerate}

\subsection{System Overview}
\begin{figure*}
    \centering
    \includegraphics[width=1.654\columnwidth]{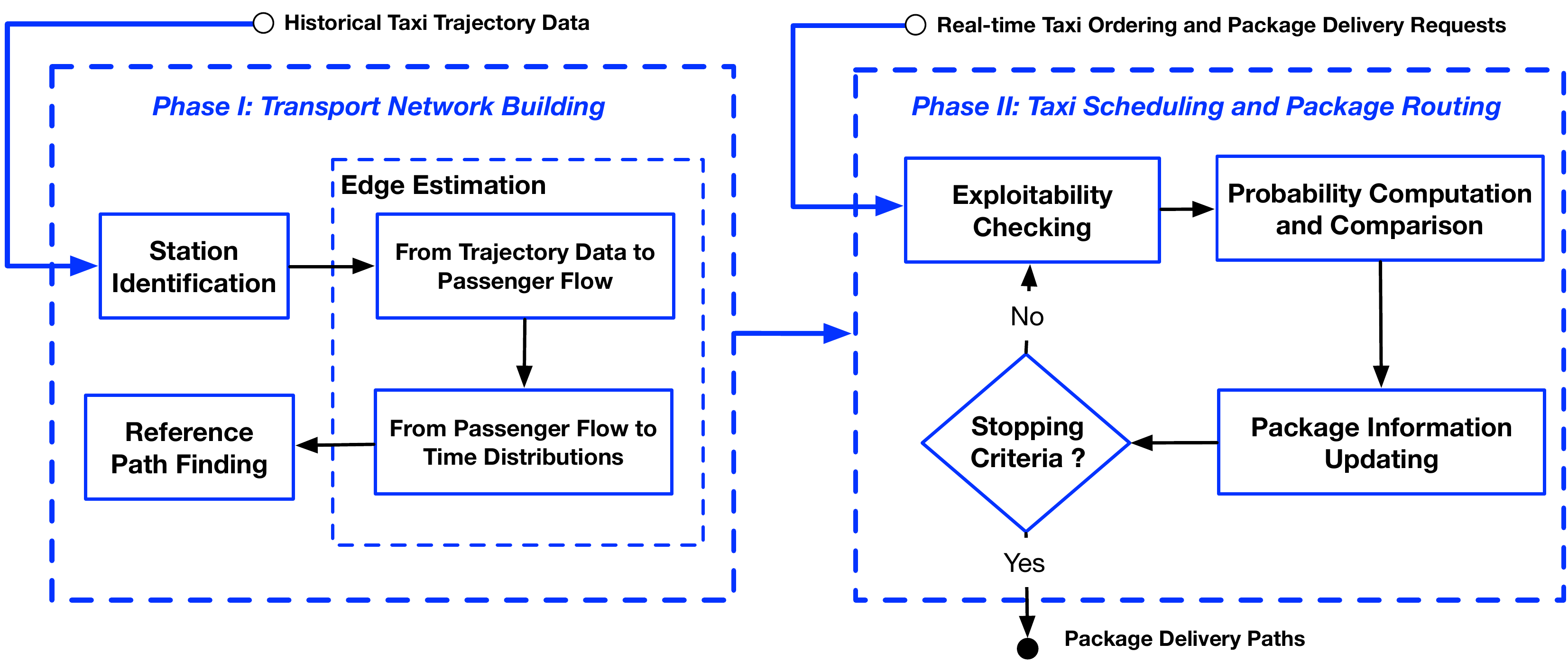}
    \caption{Overview of the CrowdExpress system.}
    \label{fig:framework}
\end{figure*}

We develop a two-phase system called CrowdExpress, i.e., {\em offline package transport network building} and {\em online taxi scheduling and package routing} to find the optimal route with the maximum probability of arriving-on-time for each package delivery request within a given deadline, by collaboratively recruiting taxi drivers that have been reserved to passengers (occupied by passengers), as shown in Fig.~\ref{fig:framework}. %with the technical details presented as follows. 

Phase I is an {\em offline}  process, with  the historical taxi trajectory data as input, aiming to identify the package interchange stations, estimate the edge values, as well as find the reference paths for any  given OD package pairs. Based on the constructed package transport network, for a real-time incoming individual package delivery request, Phase II mainly takes four {\em online} steps to tackle with, namely, {\em Exploitability Checking, Probability Computation and Comparison, Package Information Updating and Stopping Criteria Checking}, with the streaming taxi ordering requests as input. The system finally outputs the corresponding package delivery paths.  The technical details will be presented in the next two sections.

\section{Phase I: Offline Package Transport Network Building}\label{sec:phase1}
The task of offline package transport network building is to identify interchange stations (i.e., node locations) as well as the estimation of travel time distributions (i.e., edge values). Here, we mainly take a three-step procedure to achieve the objectives, detailed as follows.
\subsection{Package Interchange Station Identification}
%{\noindent}$\bullet$ {\bf Step 1: Package Interchange Station Identification.}  

\begin{figure}
\centering
\includegraphics[width=0.884\columnwidth]{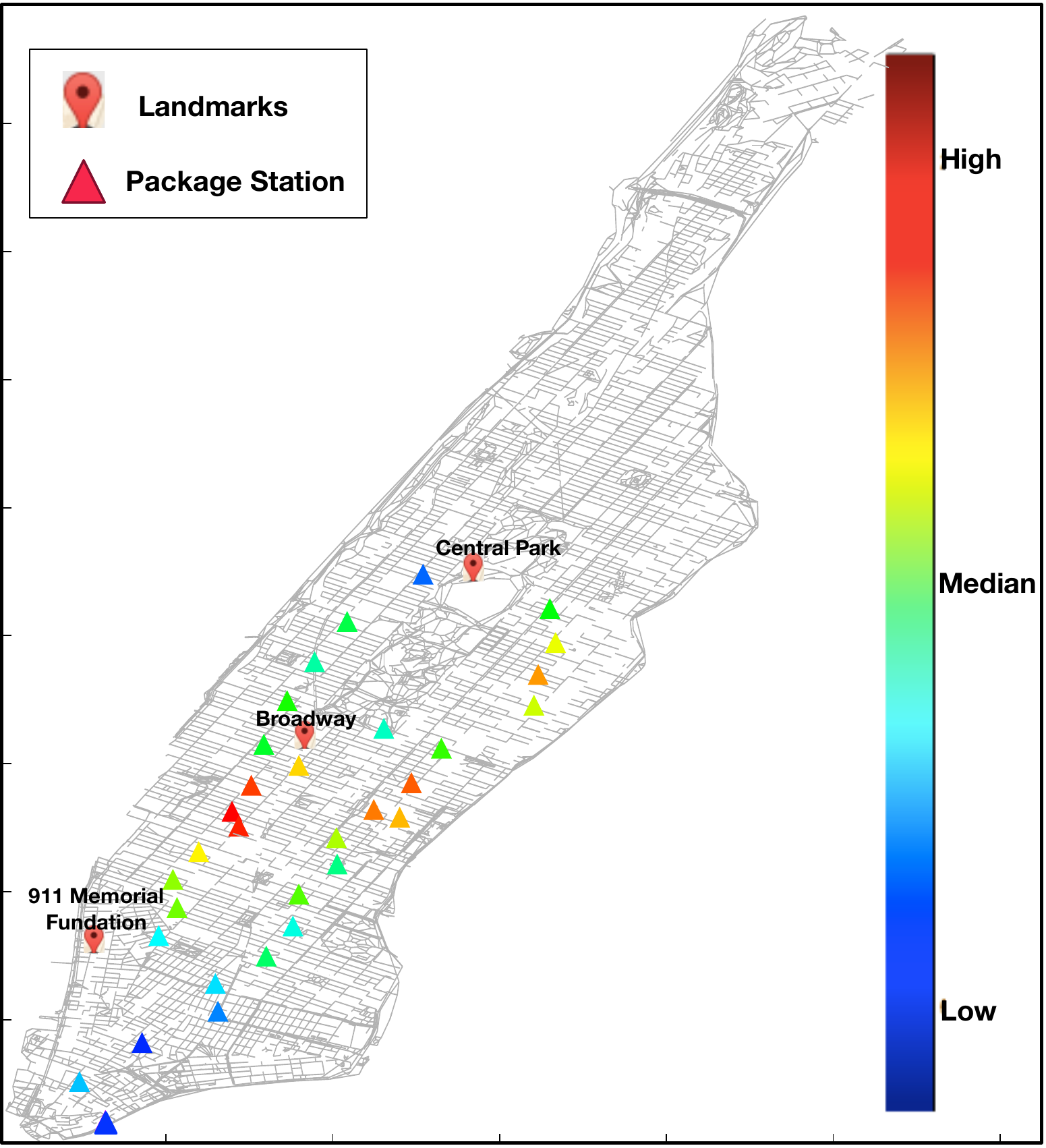}
\caption{Results of the  package interchange station  in Manhattan, NYC. The color of the station corresponds to its popularity, which is measured by the total times of the pick-ups and drop-offs.}
\label{fig:packagestations} 
\end{figure}

One basic principle for the identification of interchange stations is that they should be located where  passengers are frequently picked-up or dropped-off, to take as full advantage of passenger-sending rides for the package hitchhiking as possible, and probably to minimize the extra efforts imposed to the taxi drivers as well. Fortunately, with the taxi trajectory data left, such information (i.e., pick-up and drop-off points) can be easily extracted. Then, we cluster them using {\em DBSCAN} algorithm as it is capable of merging closer data points with arbitrary distributions~\cite{ester1996density}. Finally, locations near the point centroids of each cluster and the road sides are identified as the locations of the interchange stations to serve the purposes of  package collection, storage and receiving.  The locations of the package interchange stations in the Manhattan area of NYC is shown in Fig.~\ref{fig:packagestations}.  No package interchange stations are identified in the Upper Manhattan since there were fewer passengers take taxi to reach there.

\subsection{Edge Estimation}
\subsubsection{From Trajectory Data to Passenger Flow}
It is straightforward to infer the passenger flow between any two interchange stations during a given time slot from the trajectory data~\cite{chen16crowddeliver}. Specifically, we {\em first} group the trajectories according to their starting time ($t_s$). {\em Second}, to compute the passenger flow from $s_i$ to $s_j$, we count the number of the trajectories satisfying Eqns.~\ref{eq:pf1}$\sim$\ref{eq:pf2}.  It should be noted that there could be no passenger flow between some interchange station pairs. %Prior research has observed regular spatial and temporal patterns in human trajectory data (including taxi GPS traces), and consequently utilized these patterns in various applications~\cite{chawla2012inferring,gonzalez2008understanding,yue2009mining}. % {\em Finally}, the passenger flow from $cs_i$ to $cs_j$ is just the number of trajectories satisfying the requirements.  It should be noted that there could be no passenger flow between some collection station pairs. %of trajectories. Trajectories with the starting time falling into the same time slot will be aggregated.
\begin{equation}
Ddist(Tr_i.o,loc(s_i))\le\delta
\label{eq:pf1}
\end{equation}  
\begin{equation}
Ddist(Tr_i.d,loc(s_j))\le\delta
\label{eq:pf2}
\end{equation}  
where $Tr_i.o$ and $Tr_i.d$ are the original and destination points of $Tr_i$, respectively; $loc(\cdot)$ gets the latitude and longitude location of the given interchange station; $Ddist(a\cdot b)$ calculates the {\em driving distance} from point $a$ to point $b$; $\delta$ is a user-specified parameter.  The physical meaning of $\delta$ is that any passenger-delivery ride which starts and ends near a pair of interchange stations (i.e., with driving distance less than $\delta$) can be hitchhiked for the package delivery between this pair. Hence, for a given OD pair, a bigger $\delta$ would result in a bigger number of passenger flow. It is worth noting that, for a specific trajectory, there could be multiple interchange station pairs that satisfy  Eqns~\ref{eq:pf1}$\sim$\ref{eq:pf2}, in other words, can provide package hitchhiking ride  between all these pairs. Therefore, a bigger $\delta$  also leads to a bigger number of interchange station pairs, suggesting that the corresponding  trajectory can be more capable of providing hitchhiking rides.  However, for passengers, a bigger $\delta$ may mean a longer waiting time for the reserved taxis, since the taxi driver might have to travel farther to collect the package  before picking up passengers.  To control for the additional waiting time, we set $\delta$ to 500 meters. % (refer to Appendix~\ref{sec:appA} for more discussion). %More details can be found in Appendix. In other word, the passenger flow is relatively stronger when $\delta$ is bigger.

\subsubsection{From Passenger Flow to Time Distributions}
%{\noindent}$\bullet$ {\bf Step 3: From Passenger Flow to Time Distributions.}  

 To estimate an edge value, we need to estimate two parts, i.e. the {\em waiting time} and the {\em driving time}~\cite{chen16crowddeliver}.  The {\em driving time} is simply the travel time of each taxi trajectory.  
% the average over all such rides, which can be computed according to Eq.~\ref{eq:dt}.
%\begin{equation}
%t_d = driving~time = \frac{\sum_{i=1}^{N}Tr_i.(te-ts)}{N}
%\label{eq:dt}
%\end{equation} 
%where $N$ is the number of passenger-delivery rides during the target time slot in the observed days. $(te-ts)$ is the time cost of the corresponding taxi ride.

The {\em waiting time} on the edge is defined as the time required to wait for a suitable hitchhiking ride that can transport  a package from $s_i$ to $s_j$ directly. To address this problem, we employ the {\em Non-Homogeneous Poisson Process} (NHPP) to model the behavior of passenger taking taxis~\cite{zheng12where}. According to the statistical  {\em frequency} of passenger taking taxis from $s_i$ to $s_j$ in history (i.e. passenger flow), we can estimate the {\em waiting time} of packages at different time slots at the interchange stations. Specifically, the {\em waiting time} on the edge from $s_i$ to $s_j$ is: 
\begin{equation}
t_w = waiting~time = \frac{\Delta T}{\bar{N}}
\label{eq:wt1}
 \end{equation}
where $\bar{N}$ is the average number of passengers taking taxis from $s_i$ to $s_j$ during the given time slots; $\Delta T$ is the length of  that time slot.  Note that the {\em waiting time} obtained by Eq.~\ref{eq:wt1}  is in the {\em statistical sense}, and it could be much smaller in the real case due to the timely {\em availability} of right passenger-sending trips. % For example, $\Delta T = 9$ hours for the  day-time time slot of work days. 
 
%Finally, the edge value is  the sum of {\em waiting time} and {\em driving time} on the respected edge, as shown in Eq.~\ref{eq:tc}.
%
%\begin{equation}
%t_s = \frac{\sum_{i=1}^{N}Tr_i.(te-ts)}{N} + \frac{\Delta T}{\bar{N}}
%\label{eq:tc}
%\end{equation} 

\begin{figure}
\centering
\includegraphics[width=0.94\columnwidth]{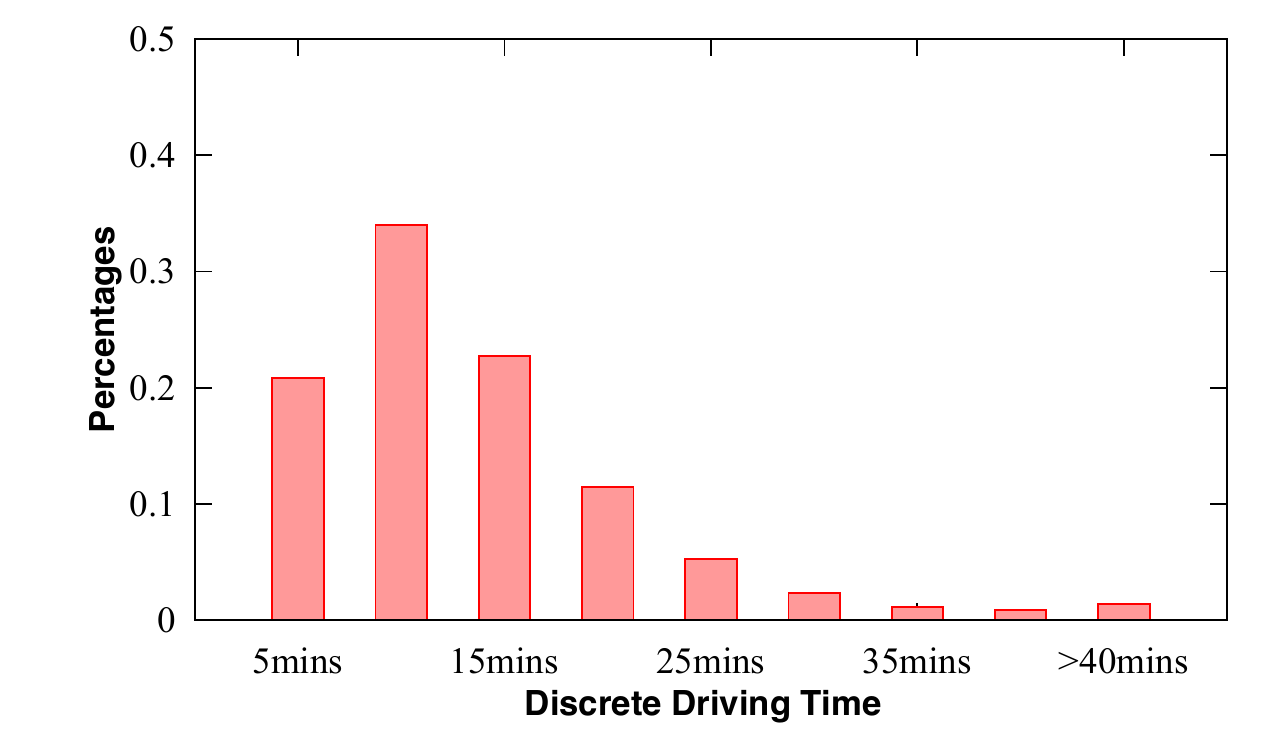}
\caption{Histogram of all driving times after discretization.}
\label{fig:drivingtime} 
\end{figure}

The waiting time component is substituted in advance when computing the arriving-on-time probability of a given path at a given time duration, thus  the corresponding {\em time distributions}  is obtained by discretizing the driving time component {\em only}  according to {\em Definition}~\ref{def:dis}. Fig.~\ref{fig:drivingtime} shows the  time distributions after discretization. As one can observe,   driving times greater than 40 minutes occupy quite a small percentage, i.e., less than 1.38\% in the taxi trajectory data in  NYC.  Note that the edge value would be  $+\infty$ if there was no passenger flow (or less than a certain amount) on the respected edge in history.

\subsection{Reference Path Finding}
%{\noindent}$\bullet$ {\bf  Step 4 Reference Path Finding.}  

On the basis of the constructed package transport network, we find two reference paths for each given OD pair, i.e., the shortest path when assuming value on each edge is the minimum travel time ($SPath\_min$), and the shortest one when assuming value on each edge is the maximum travel time ($SPath\_max$), respectively. More specifically, given an OD pair, when choosing all minimum travel times on  edges,  we can find the shortest path, i.e. $SPath\_min$,  using {\em Dijkstra's} algorithm~\cite{skiena1990dijkstra}. $SPath\_min$ refers to the case that the package can be  delivered  in the most efficient manner if  sent via that path.  Similarly, when choosing all maximum travel times on  edges, we can also discover the shortest path, i.e. $SPath\_max$. $SPath\_max$, as a comparison,  refers to the case that the package can be delivered in the lowest efficient manner if  sent via that path. Moreover, the corresponding total travel times of those two paths are also recorded, which can be used to guide the online taxi scheduling and  package routing upon the real-time taxi ordering requests  in  the second phase, with details would be further addressed in the next section. It should be noted that, although it is a time-consuming procedure to find two shortest paths for each  OD pair with a computation complexity of $\mathcal{O}(k^2)$, it can be operated offline\footnote{Suppose there are $k$ nodes in the network, the maximum number of all  OD pairs is $k(k-1)$.}. 
 
\section{Phase II: Online Taxi Scheduling and Package Routing}\label{sec:phase2}
Given an OD pair of the package, the task of {\em online taxi scheduling} is the decision-making. To be more specific, the phase should make a decision whether utilizing the {\em current} available  hitchhiking ride for the package delivery or waiting for the  {\em future} rides, according to the upcoming taxi ordering requests generated in real-time and the remaining time margin.  While the task of {\em online package routing} is much simpler, i.e., just assigning the delivery task to the scheduled taxi. As a result,  the package will be delivered to the next stop, which is same to the destination of the taxi. The two steps impact each other {\em mutually}. On one hand, both the potential hitchhiking rides for the package delivery and the time margin is highly related to the current stop where the package locates; on the other hand, which the next stop that the package will head to is determined by the scheduled taxi. In the following, we mainly focus on the  step of online taxi scheduling, which includes the following operations.% the pseudo-code of which is illustrated in Algorithm~.

\subsection{Exploitability Checking}
Before triggering this phase, we first need to conduct the {\em exploitability checking}, i.e., to determine: 1) whether the origin of an upcoming taxi ordering request is close to the location of the package; and 2) whether it ends at one of the interchange stations. If both conditions are met, then we further need to compare the maximum arriving-on-time probability of sending the package via $s_k$ ({\bf hitchhiking the current}) to the maximum arriving-on-time probability of sending the package via other potential next stations (the other neighbours of $s_o$ in the transport network) ({\bf waiting for the future}).   If the former value is greater, then the package will be sent out {\em immediately} by hitchhiking the {\em current} taxi ride; otherwise, the system will {\em wait} for the new {\em future} taxi ordering requests that may lead to a  higher arriving-on-time probability and the decision will made again, given the new time margin. Thus, the core of the online taxi scheduling is the maximum arriving-on-time probabilities {\em computation and comparison}. 

\subsection{Probability Computation and Comparison}
\subsubsection{Probability Computation if Hitchhiking the Current}
%{\noindent}$\bullet$ {\bf Step 1: Probability Computation.}
%
%$\bullet$ {\bf Step 1.1: Probability Computation if Hitchhiking the Current.}

According to the definition, the maximum arriving-on-time probability for case {\em if hitchhiking  the current} (suppose that the recruited taxi will go to $s_k$) can be computed as follows:
\begin{equation}
    P_{od}(t\le t_0-\Delta t|Path_{okd}) = \int\limits_{0}^{t_0-\Delta t}P_{ok}(\omega)u_{kd}(t\le t_0-\Delta t)d\omega
    \label{eq:maximum}
\end{equation}
where $Path_{okd}$ refers to the path from $s_o$ to $s_d$ while stopping at  $s_k$ in the next; $\Delta t$ is the time difference between the occurring time of the taxi ordering request and the birth time of the package delivery request, i.e. $\Delta t = tor.t-pr.t$; $t_0$ is the given time duration of the deadline; $u_{kd}$ is the maximum arriving-on-time probability from $s_k$ to $s_d$, as defined in {\em Definition}~\ref{def:maximum}.

By definition, it is easy to compute the arriving-on-time probability of a determined path under a given deadline time, such as $\int\nolimits_{0}^{t_0-\Delta t}P_{ok}(t)dt$. By contrast, it is rather challenging to get the value of the latter part in Eq.~\ref{eq:maximum}. As discussed, one naive and straightforward way is {\em first}  to enumerate all the possible paths from $s_k$ to $s_d$, {\em then} compute the value of arriving-on-time probability for each of them, {\em finally} pick up the maximum one as the final value. It is easy to understand that the trivial method cannot work in real cases as the problem of finding all possible path for a given OD pair is NP-hard. Actually, it is also no need and some branches in the transport network can be {\em trimmed recursively}. We propose a novel algorithm named {\bf maxProb} to compute the probability, which mainly consists of two operations, i.e, initialization and deep-first searching.%, which will be detailed as follows.   %We use the example shown in Fig. to illustrate the idea.
%{\small%\hspace{-1.2em}
%\fbox{%
%\parbox{0.98\columnwidth}{%

{\color{black}
{\em Initialization}: From $s_k$ to $s_d$, it will be easy to find two shortest paths, $SPath\_min(s_k,s_d)$, $SPath\_max(s_k,s_d)$. We further obtain the two corresponding reference paths from $s_o$ to $s_d$ via $s_k$,  and compute their arriving-on-time probabilities given the remaining time, which are two boundaries and used to guide the process of  {\em branch trimming}. For brevity, we use $minP_{o\rightarrow k\leadsto d}$ and $maxP_{o\rightarrow k\leadsto d}$\footnote{$a\rightarrow b$ refers to the package is sent from $s_a$ to $s_b$ directly, while $a\leadsto b$ refers to the package is sent from $s_a$ to $s_b$ via some intermediate stops that are determined by the reference paths between $s_a$ and $s_b$.} to represent the arriving-on-time probabilities of $s_o\rightarrow SPath\_min(s_k,s_d)$ and $s_o\rightarrow SPath\_max(s_k,s_d)$, respectively. 
% from $s_o$ to $s_d$ via $s_{k_1}$, $s_{k_2}$ and $s_{k_i}$ sequentially, respectively\footnote{$minP$ refers to the probability of $SPath\_min$, while $maxP$ refers to probability of $SPath\_max$.}.

%Note that  $Path\_ref_1$ = $s_i+s_k+SPath\_min(s_k,s_d)$, $Path\_ref_2$ = $s_i+s_k+SPath\_max(s_k,s_d)$\footnote{$a+b$ denotes that the package first arrives at $a$ then $b$.}; $ct_1$ and $ct_2$ refer to the travel time on two reference paths respectively.   denoted as $Path\_minRef$ and $Path\_maxRef$, respectively,

\begin{algorithm}[htb]
\caption{$FindNextStop(s_k,s_d,TN)$}
\begin{algorithmic}[1]
\STATE $ngb=Neigh(s_k,TN)$; \COMMENT{get the neighbouring stations of $s_k$ based on the transport network topology.}
\STATE{$ns=\emptyset$};
\STATE{$refP=minP_{o\rightarrow k\leadsto d}$};
\FOR{$i == 1$ to $|ngb|$}
\STATE $s_{ki} = ngb(i)$;
\IF{
$refP \le \newline
P(t\le t_0-\Delta t-t\_min(s_{ki},s_d)|Path_{o\rightarrow k \rightarrow ki})$
}
\STATE $ns=ns\cup ngb(i)$;
\ENDIF
\ENDFOR
%\STATE $\tau_c=\tau_c \cup e_{|\tau_m|}$;
\end{algorithmic}
\label{alg:maxprob}
\end{algorithm}

{\em Depth-First-Searching}: From $s_k$ to $s_d$, we mainly apply the {\em Depth-First-Search} ({\em DFS}) method to recursively get each possible path~\cite{tarjan1972depth}, and compute the maximum probability of arriving-on-time. One exception is that the user specifies an extremely long deadline, mathematically, $maxP_{o\rightarrow k\leadsto d}=1$, implying that the package can be delivered on time for sure via the reference path. Therefore, no {\em DFS} is needed and a simple taxi scheduling can be enough under such circumstance. The overall procedure of {\em DFS} starting from $s_k$ can be summarized as follows. 

 The core function is to find the next package stop of $s_k$, with the pseudocode shown in Algorithm~\ref{alg:maxprob}. The very beginning task is to get the neighouring stations of $s_k$, given the topology of the built package transport network (Line 1).   {\em DFS} starts to find the next stop from one of the neighbouring nodes of $s_k$ (e.g., $s_{ki}$)  in the loop (Lines 4$\sim$9). Whether $s_{ki}$ can be the next package stop is determined by the  inequation shown in Line 6. In the in-equation, the reference probability ($refP$) is first set to $minP_{o\rightarrow k\leadsto d}$ (Line 3).  $t\_min(s_{ki},s_d)$ is the time cost of the reference path $SPath\_min(s_{ki},s_d)$ estimated by the historical taxi trajectory data; the right part of the in-equation is the maximum arriving-on-time probability which corresponds to the {\em ideal} case that the package can be shipped from $s_{ki}$ to $s_d$ in the most-efficient way (i.e., the time cost on each edge  in the package transport network is the minimal). If the in-equation satisfies,  {\em it indicates that there exists a potential path which can lead to a higher maximum arriving-on-time probability than the reference path}, thus {\em DFS} will continue to search with a new start from $s_{ki}$ {\em recursively}, with the same procedure to the {\em DFS} starting from $s_k$; otherwise, {\em DFS} will be terminated and the related branches will be trimmed at the same time. Thus, a recursion may be stopped either at some intermediate node or generates a successful path reaching the given destination. If a {\em valid} path is resulted ($Path\_valid$), $refP$ will be updated using its corresponding probability {\em if and only if} it is greater than the previous value.

%\begin{equation}
%refP \le P(t\le t_0-\Delta t-t\_min(s_{ki},s_d)|Path_{o\rightarrow k \rightarrow ki})
%    \label{eq:dfs}
%\end{equation}
%where 

The whole {\em DFS} ends when all neighhours of $s_k$ are checked by {\em repeatedly} calling the above recursive {\em DFS}. Finally, the maximum arriving-on-time probability if assigning the package delivery task to the current available taxi shall be the final value of $refP$. 
}
 %and the  will be conducted. The recursion will be stopped either at some intermediate node or a successful path reaching the given destination is generated. For the latter case, $refPath$    

%{\em DFS} will pick up any one of the neighouring nodes of $s_{ki}$ and continue to find the new path. 
%
%
%For each neighbouring nodes of $s_k$, at the first depth,   
%
%
%Specifically, we apply {\em DFS} for each of neigbhouring nodes of $s_k$. {\em DFS} will be terminated at some {\em intermediate} node (not the destination) and the related branches will be trimmed at the same time, if all package-delivery paths that stop at that node cannot achieve a higher arriving-on-time probability than $minP$. Otherwise, a new path from $s_k$ to $s_d$ will be discovered, and $minP$ will be updated by its corresponding arriving-on-time probability for the future {\em DFS}.  The maximum arriving-on-time probability if assigning the package delivery task to the current available taxi will be the value of final $minP$. %We use the example shown in Fig to illustrate the above-mentioned procedure. %
%  }%
%}}

%$\bullet$ {\bf Step 1.2: Probability Computation if Waiting for the Future.}
\subsubsection{Probability Computation if Waiting for the Future}
If the future taxi rides heading to any one of the other neighouring nodes of $s_o$ except for $s_k$ (marked as $\{ngb(s_o)\}-s_k$) could lead to a higher arriving-on-time probability, compared to the case if {\em hitchhiking the current}, a better decision should be the {\em waiting}. The maximum arriving-on-time probability {\em if waiting for the future}  can be computed as follows:
\begin{equation}
\begin{aligned}
    P_{od}(t\le t_0-\Delta t'|Path_{ojd}) =\\
     \max_{s_j\in \{ngb(s_o)\}-s_k }\int\limits_{0}^{t_0-\Delta t'}P_{oj}(\omega)u_{jd}(t\le t_0-\Delta t')d\omega
\end{aligned}
\label{eq:maximumW}
\end{equation}
where $\Delta t'=\Delta t+t_w(s_o,s_j)$ and $t_w(s_o,s_j)$ refers to the edge value component of waiting time from $s_o$ to $s_j$. As can be seen, the major difference between Eq.~\ref{eq:maximum} and Eq.~\ref{eq:maximumW} is the time margin. More specifically, less time margin is left for the package deliveries as  an additional time cost would be induced while waiting for the future taxi rides. Here, we simply use the average waiting time to approximate the additional time cost.  

Similarly, all maximum arriving-on-time probabilities of waiting for the future exploitable taxi rides from $s_o$ can be computed, and the maximal one among them will be chosen to represent the maximum arriving-on-time probability if assigning the package delivery task to the future taxis. 
    
\subsubsection{Probability Comparison}
%{\noindent}$\bullet$ {\bf Step 2: Probability Comparison.}

As discussed, once receiving a real-time taxi ordering request, {\em on-line taxi scheduling and package routing} will be activated, and the package may be shipped to some intermediate stop by hitchhiking the current ride or stands still at the current stop by comparing those two maximum arriving-on-time probabilities. Note that the remaining time margin shrinks as time goes by,  the two probabilities computed in Eqns.~\ref{eq:maximum} and \ref{eq:maximumW} are {\em dynamically changed}, thus the better decision (hitchhiking the current or waiting for the future) can be also adjusted {\em adaptively} ``on-the-fly''. 

\subsection{Package Information Updating}
After the package is sent to the next station whether by hitchhiking the current or future rides, the information about the package delivery request should be updated. {\color{black}To be more specific, the origin of the package should be set as the updated station that the package locates;} the birth time should be set as the time when the package arrives at the current stop. The newly updated package delivery request will be used as the input of Phase II.  

\subsection{Stopping Criteria Check}
For a package delivery, the previous three operations  will be {\em iteratively} conducted until one of the following two stopping criteria is satisfied: 1) the package has arrived at its destination; 2) the time is running out (the package cannot be delivered by the deadline), in that case, the system would report {\em failure}. For those failure package deliveries, empty taxis can be recruited dedicatedly to send them to the destinations. However, the topic is out of the scope of the paper.  

\section{Experimental Evaluations}\label{sec:evaluation}
In this section, we empirically evaluate  the performance of the proposed {\bf maxProb} algorithm. We first introduce the experimental setup, baseline algorithms used for comparison, evaluation metrics and results on algorithm efficiency and effectiveness.  We discuss some open research issues to be further addressed  in the end.  

\subsection{Experimental Setup}
\subsubsection{Experimental Data} We  use  the real-world datasets for the evaluation, i.e. the road network data which is extracted from OpenStreetMap\footnote{\url{www.openstreetmap.org}}, and one month of taxi trajectory data generated by over 19,000 taxis in the city of New York (NYC), US. Readers can refer to~\cite{alarabi2014tareeg} for the details on how to extract the road network from the crowd-sourced open platform (i.e., OpenStreetMap) correctly.  We determine package interchange stations according to the  algorithm discussed earlier.

 For the taxi trajectory data, we split it data into training and testing sets, according to the date of the month. Specifically, the training set contains taxi trajectories on 1st$\sim$20th, January, 2013, which are used to build package transport network. It should be noted that for the taxi trajectory data in NYC, no detailed travel routes between the pick-up and drop-off points are provided due to the privacy considerations. The testing  trajectories were  generated from 21st to 31st,  January, 2013, which are used as the real-time taxi ordering requests ($TOR$) for testing the performance of the proposed {\bf maxProb} algorithm.  Table~\ref{tab:statistics} shows some basic statistics of the taxi  trajectory  and road network data.

\begin{table}
\centering
\caption{Statistics of the taxi trajectory and road network  data sets.}
\label{tab:statistics}
\begin{tabular}{c||c|c}
\toprule
Datasets       & Properties               & Statistics \\
\hline\hline
Taxi  trajectory & Number of taxis          &   >19,000         \\
                & Number of occupied rides &     $\approx$ 13 M      \\
                \hline
Road network     & Number of  road intersections         &     11,999       \\
                & Number of road segments          &     15,202      \\
\bottomrule
\end{tabular}
\vspace{-0.4cm}
\end{table}

\subsubsection{Package Delivery Request} {\color{black} Since the data sets do not contain information about package delivery requests, we apply simple mechanism to simulate it. The novel package delivery system targets the city-wide person-to-person service. Hence, to simulate a package delivery request ($PDR$), we randomly generate its birth time, origin and  destination.  Regarding the origin and destination, any package should be originated and ended at the interchange stations.}  %We further eliminate  requests with short-distance OD pairs since few users would request speedy shipping as ordinary delivery may be equally efficient in this case.

\subsubsection{Evaluation Environment}
All the evaluations in the paper are programmed using Java language under the Eclipse J2SE 1.5 integrated development environment,  and run  on an Intel Core i5-4950 PC with 8-GB RAM and Windows 7 operation system.

\subsubsection{Evaluation Metrics} We adopt the following three metrics to evaluate the proposed {\bf maxProb}.

{\bf Success Rate.} The success rate is the ratio of the number of packages which can be delivered successfully within a given deadline (i.e. time duration) to the number of total packages (i.e. the number of package delivery requests simulated).
\begin{equation}
SR(t\le deadT) = \frac{|\mathcal{TC}(Path(o_p\leadsto d_p))\le deadT|}{|PQ|}
\end{equation} 
where $Path(o_p\leadsto d_p)$ represents the  optimal  path  generated by the proposed {\bf maxProb} algorithm for a given package delivery request; $deadT$ is the given deadline.  The delivery performance is better if the success rate is higher within a given shorter deadline. %This metric also takes into  consideration  the package delivery time collectively.

Regarding the {\em deadline setting},  we do not set an {\em absolute} value for all package deliveries since package originated (ended) at different locations would need {\em absolutely} varied time. Thus, for an individual package delivery, we set a {\em relative} deadline separately instead, according to the following equation:
\begin{equation}
deadT = t_{avg} + extraT    \label{eq:deadlines}
\end{equation}
in which $t_{avg}$ is the  average  value of the time cost by the two reference paths, which is obviously different for  packages with different OD pairs.  $extraT$ is the  extra time value imposed by the user; a smaller $extraT$ indicates that the user needs the package more urgently and wants it to be arrived more timely.

{\bf Number of Relays.} The number of relays ($Num_{relays}$) during a package delivery  is  defined as the number  of participating taxis $Num_{p\_taxis}$ (Formula~\ref{eq:relays}).
\begin{equation}
Num_{relays} = Num_{p\_taxis} 
\label{eq:relays}
\end{equation}
 On one hand, fewer relays generally mean a lower chance of  package loss or damage, and perhaps less overhead cost.  On the other hand, fewer number of participating taxis may imply requiring less reward cost to taxi drivers.  Thus, the performance is better if the number of relays is smaller. 
   
{\bf Package Throughput.} The average number of package deliveries that the system can complete successfully per day. The system achieves better performance if the package throughput is bigger. 

%\subsubsection{Baseline Algorithms}
%\subsubsection{Evaluation Metrics}
\subsection{Baseline Algorithms}
To show the superior performance of our proposed algorithm, we compare it with the following three baseline algorithms. \\
(1) {\bf FCFS} - This method adopts the {\em First Come First Service} strategy. Specifically, the package will be assigned to the {\em first} taxi that will pick up a passenger near the interchange station that the package locates, {\em regardless of its destination}. In fact, this algorithm always favors the strategy of {\em hitchhiking the current}, which is also known as an extension of the simple and well-known {\em flooding} strategy~\cite{Balasubramanian07DRR,Ko02flooding}.\\
(2) {\bf DesCloser} - This method assigns the package to the {\em first} taxi that will head to somewhere closer to the destination of the package, compared to the current station of the package. This algorithm  implements a distance-based geo-cast scheme that is commonly seen in other domains~\cite{zhang14geomob,zorzi03geraf}. \\
(3) {\bf Direct} - This method waits for the taxi heading to the destination near the interchange stations that the package will be delivered {\em directly}, without any intermediate stops. Specifically, the package will be assigned to the taxi that will pick up  and drop off a passenger near the interchange stations that the package locates and heads to, respectively. Thus, no relays are needed.

{\bf Remark.} Each relay in {\bf DesCloser} is effective as it ensures that the package would move towards its destination step by step; while some relays in {\bf FCFS} can be   ineffective as the package  moves further away from  its destination.  For {\bf Direct}, it may be inefficient for package deliveries where there is little passenger flow in-between. However, all three baseline algorithms do not take the arriving-on-time probability of package deliveries into account, thus probably resulting in a high failure rate.

\subsection{Experimental Objectives}
We plan the experiments  to  address the following questions. %The efficiency is measured by the time needed to response an incoming package delivery query, and the effectiveness is measured by the metrics defined in Section, including {\em Success Rate} and {\em Number of Transshipments}. 

{\em Question 1: How much computational resource is required to generate the response for a package delivery request? } 

{\em Question 2:  How does {\bf maxProb} perform  under different given deadlines? }

{\em Question 3: How does {\bf maxProb} perform  w.r.t the  birth time of package deliveries?} %\item {\em Q2:} Is there OD pairs that commonly failed to be delivered? % investigate the root cause % in terms of the success rate and the number of relays

{\em Question 4: How does {\bf maxProb} perform  w.r.t the  locations (both origins and destinations) of package deliveries?} %\item {\em Q2:} Is there OD pairs that commonly failed to be delivered? % investigate the root cause %in terms of the success rate and the number of relays 

%{\em Question 4: How far the  solution obtained by the proposed algorithm is from the optimal one?}

{\em Question 5: How many packages can be delivered daily on average (i.e.,   throughput) with the proposed system?}

The first question concerns the efficiency  of  {\bf maxProb}, and {\em Questions} 2$\sim$4 are related to its effectiveness. To answer the first question, we compute the {\em response time} of the algorithms. Since passenger flows are both time- and space-dependent, to assess the effectiveness of the different algorithms, we calculate their success rates and the number of relays with respect to packages to be dispatched to different parts of the city at different time of the day. We test the throughput  of the proposed system and the success rate under different number of package requests generated per hour, and also examine the system throughput given  different number (density) of interchange stations in the designate city ({\em Question 5}).

\subsection{Experimental Results}
\subsubsection{Results of Response Time}
We first analyze the main {\em operations} involved in the four algorithms respectively. For a given package delivery request ($pr$), when a new real-time taxi ordering request ($tor$) comes in, all four algorithms need to determine  whether $tor$ starts near the origin of the package and stops at some interchange station, i.e., {\em exploitability checking}. {\bf FCFS} will recruit the first taxi that satisfies the criteria, but for {\bf DesCloser}, it needs to further determine whether the heading destination of the taxi is closer to the destination of the package, compared to its current location. For {\bf Direct}, it also needs to determine whether the taxi would head to the destination of the package. Thus one more {\em comparison} operation is required for both {\bf DesCloser} and {\bf Direct} algorithms. For {\bf maxProb}, the procedure is even more complicated, mainly requiring additional {\em probability computation and comparison} operations, as discussed previously.  Each algorithm needs to repeat its own operation procedure at each intermediate  station (except for {\bf Direct}) and thus the total response time is the {\em accumulated} computational time over all iterations. 
\begin{figure}
    \centering
    \includegraphics[width=0.894\columnwidth]{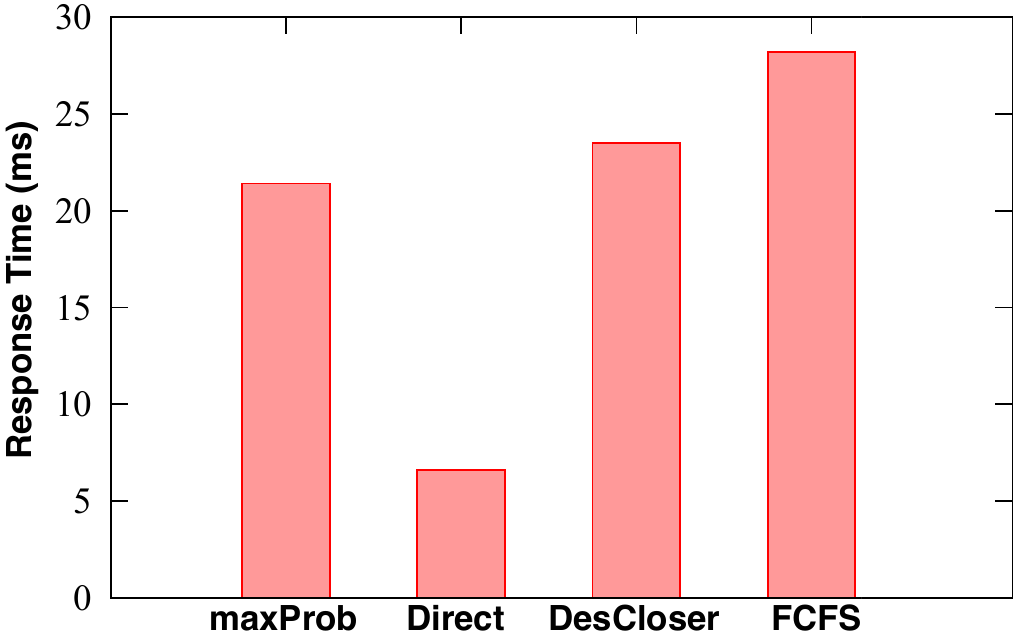}
    \caption{Evaluation results of response time for four different algorithms.}
    \label{fig:responseTime}
\end{figure}

We show the  comparison results of average response time  of the four different algorithms in Fig.~\ref{fig:responseTime}.  The average response time of {\bf FCFS} is the biggest  while that of {\bf Direct} is the smallest among all algorithms. The average response times of {\bf DesCloser} and {\bf maxProb} are in-between and {\bf DesCloser} costs slightly more time than {\bf maxProb}.  More precisely, the average response time of {\bf Direct} is within 7 milliseconds; the average response time of {\bf maxProb} is around 22 milliseconds; the  average response time of {\bf FCFS} is no more than 30 milliseconds. All results indicate that all  four algorithms are quite efficient, and can plan and adjust the shipping routes in real time. We also observe an interesting phenomenon: although {\bf FCFS} only involves two simple comparisons for each candidate taxi, it is the most time-consuming method accumulatively. In comparison,  {\bf maxProb} which contains the most sophisticated operations needs a shorter response time than {\bf FCFS} and {\bf DesCloser}. We argue that this is because: {\bf FCFS} and  {\bf DesCloser} require  more rounds of computation (i.e., more relays) than the {\bf maxProb} algorithm. %The detailed information about the simulated package delivery requests  for this study is shown in . Among the 15,000  requests, the average driving distance of the package OD pairs is 21.832 kilometers, and the driving distance can be  up to 78.453 kilometers. We randomly generated the birth time of the package delivery requests, and thus they were distributed rather evenly at the given time slots. For brevity, we do not show the request distributions on time for all experiments.

\begin{figure}
    \centering
    \includegraphics[width=0.894\columnwidth]{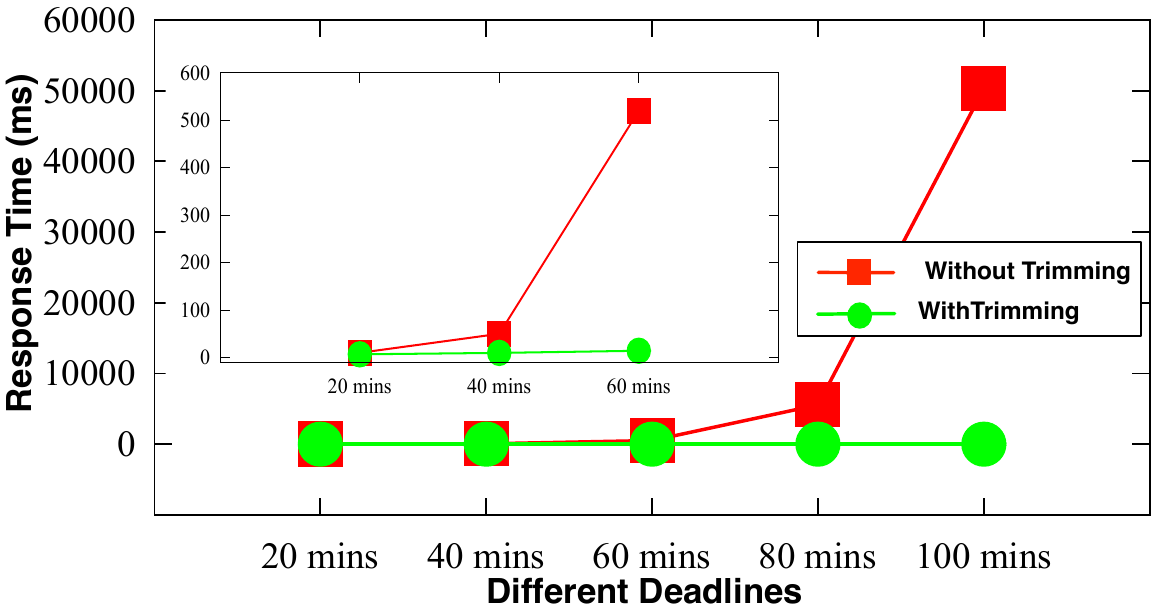}
    \caption{Comparison results of response time of {\bf maxProb} with/without branch trimming w.r.t different deadlines.}
    \label{fig:responseTimeComp}
\end{figure}

We further evaluate the effectiveness of the branch trimming in the probability computation for {\bf maxProb} in terms of the average response time saving, with the result shown in Fig.~\ref{fig:responseTimeComp}. To better illustration, we also highlight the result w.r.t different deadlines within the range of [20, 40] minutes in the left-top part of the figure.  A significant time saving is obtained with the introduction of branch trimming. To be more specific, the gap of the average response time with/without branch trimming increases exponentially wider as the given deadline becomes bigger, what is more, all the average response times with branch trimming remain stable and small under all given deadlines. The package delivery requests are the same for the efficiency studies, with a number of 100.
  
\subsubsection{Results of Success Rate}

\begin{figure}
    \centering
    \includegraphics[width=0.884\columnwidth]{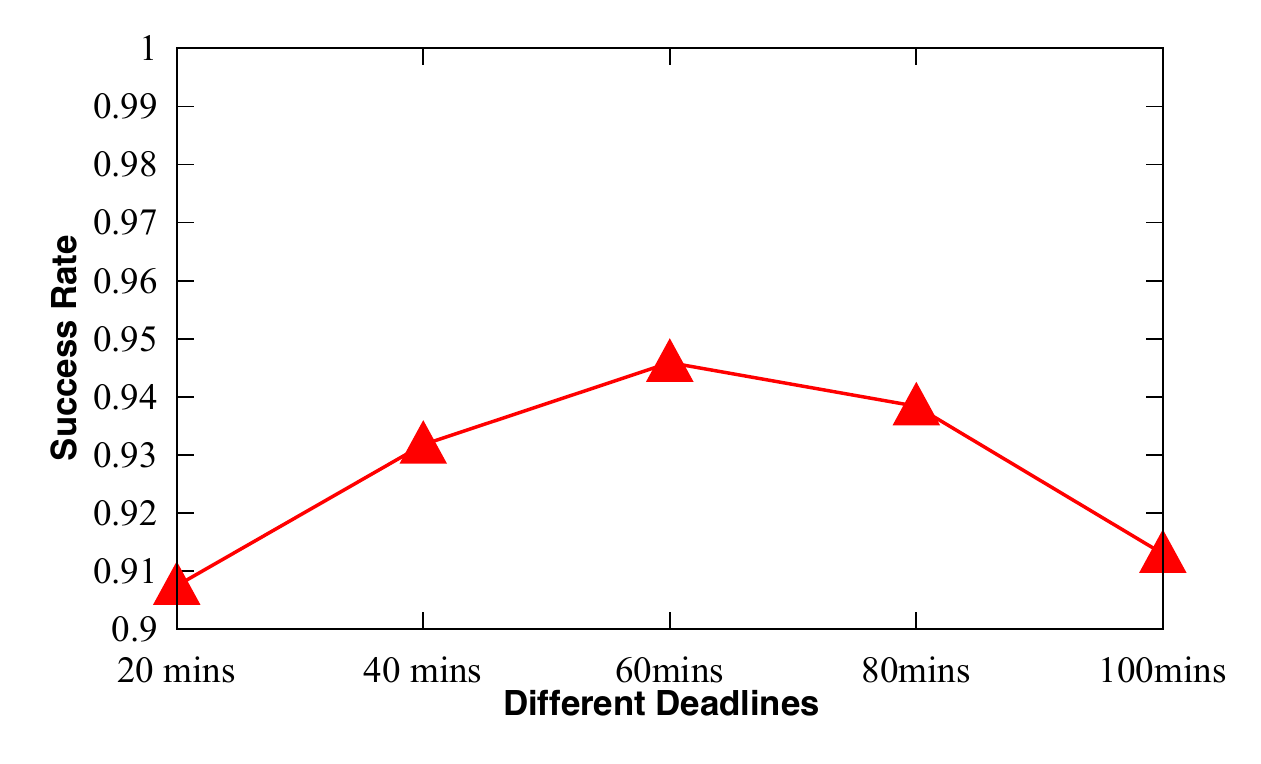}
    \caption{Evaluation results of success rate under different  deadlines.}
    \label{fig:srdeadlines}
\end{figure}
{\color{black}
We present  results of the performance of {\bf maxProb} in terms of the success rate under different deadlines in Fig.~\ref{fig:srdeadlines}. More specifically, $extraT$ is set in a range from 20 to 100 minutes, with an equal interval of 20 minutes.  As one can see, the success rate  under all deadlines is high, with a value above 90\%. The success rate firstly becomes slightly higher then decreases gradually as  users allow a longer deadline. The highest success rate appears when the $extraT$ is set 60 minutes, with the value of around 95\%. The observed phenomena seems {\em somehow counterfactual} at the first glance as the success rate should be higher when setting a longer deadline in intuition. The root cause is that: when giving a bigger deadline, the arriving-on-time probability if hitching the current becomes greater, as guaranteed by Theorem~\ref{th:biggerdeadline}. An extreme case is that the probability would always equal one and  {\em dominates} the other possibilities, as a consequence, the {\bf maxProb} algorithm tends to select the {\em hitchhiking the current} strategy  while routing packages.  Under such circumstance, {\bf maxProb} degrades to the {\bf FCFS} algorithm to some extent, causing the negligible decrease of the success rate. To get rid of such degradation, the key issue is to lower the value computed by Eqn.~\ref{eq:maximum}. Thus, one potential solution can be the reduction of remaining time margin during package routing when applying {\bf maxProb} algorithm. For instance, if {\em hitchhiking the current} strategy is always preferred during package routing, the remaining time margin can be manually reduced to 90\% of the true one that imposed by the user.  It should be noted that the success rate of {\bf FCFS} is much lower, compared to {\bf maxProb}, which will be verified in the following experiments. The number of package delivery requests is 10,000, with the birth time uniformly distributed at the day-time (i.e., from 8:00 to 18:00).
}
\begin{figure}
    \centering
    \subfigure[]{\includegraphics[width=0.884\columnwidth]{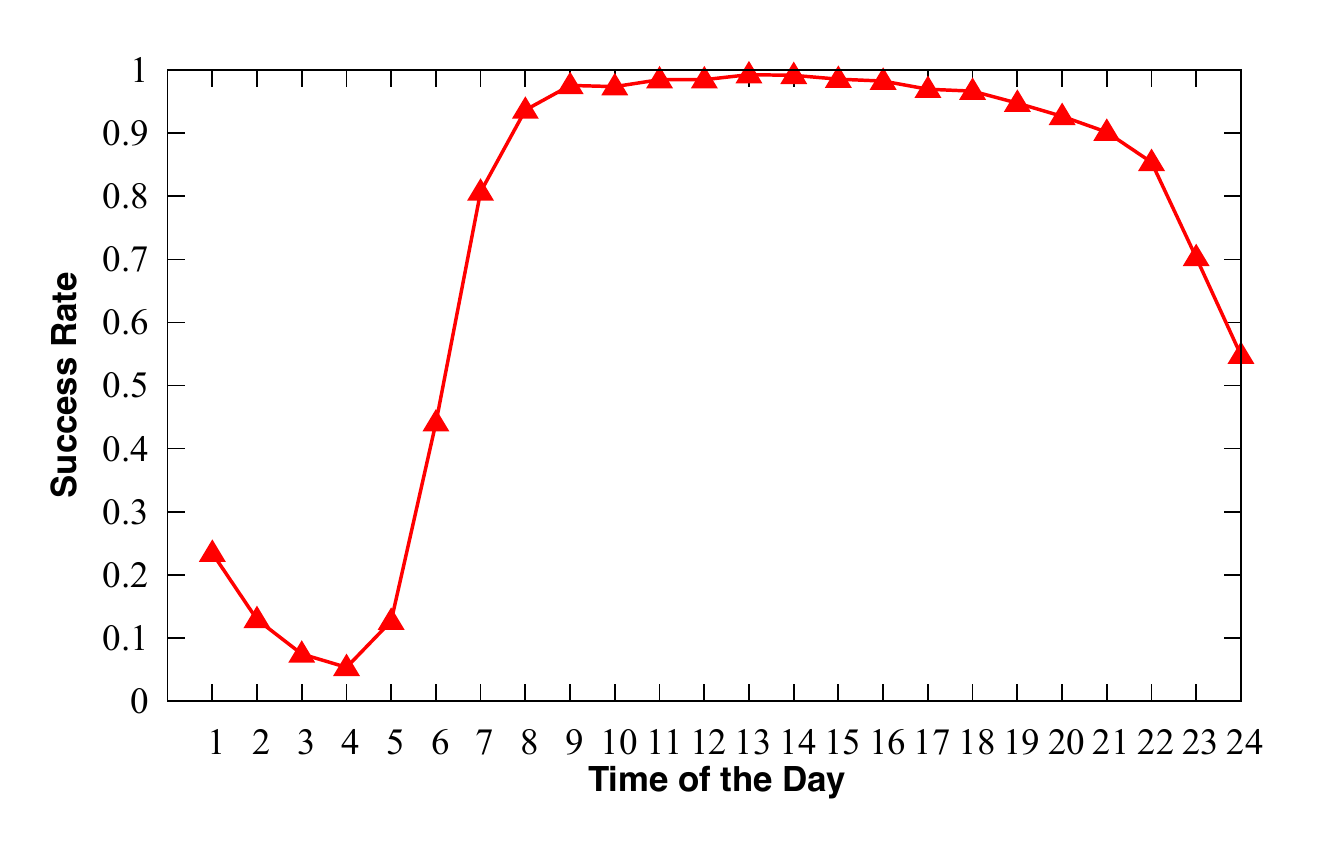}\label{fig:srhours}}
    \subfigure[]{\includegraphics[width=0.884\columnwidth]{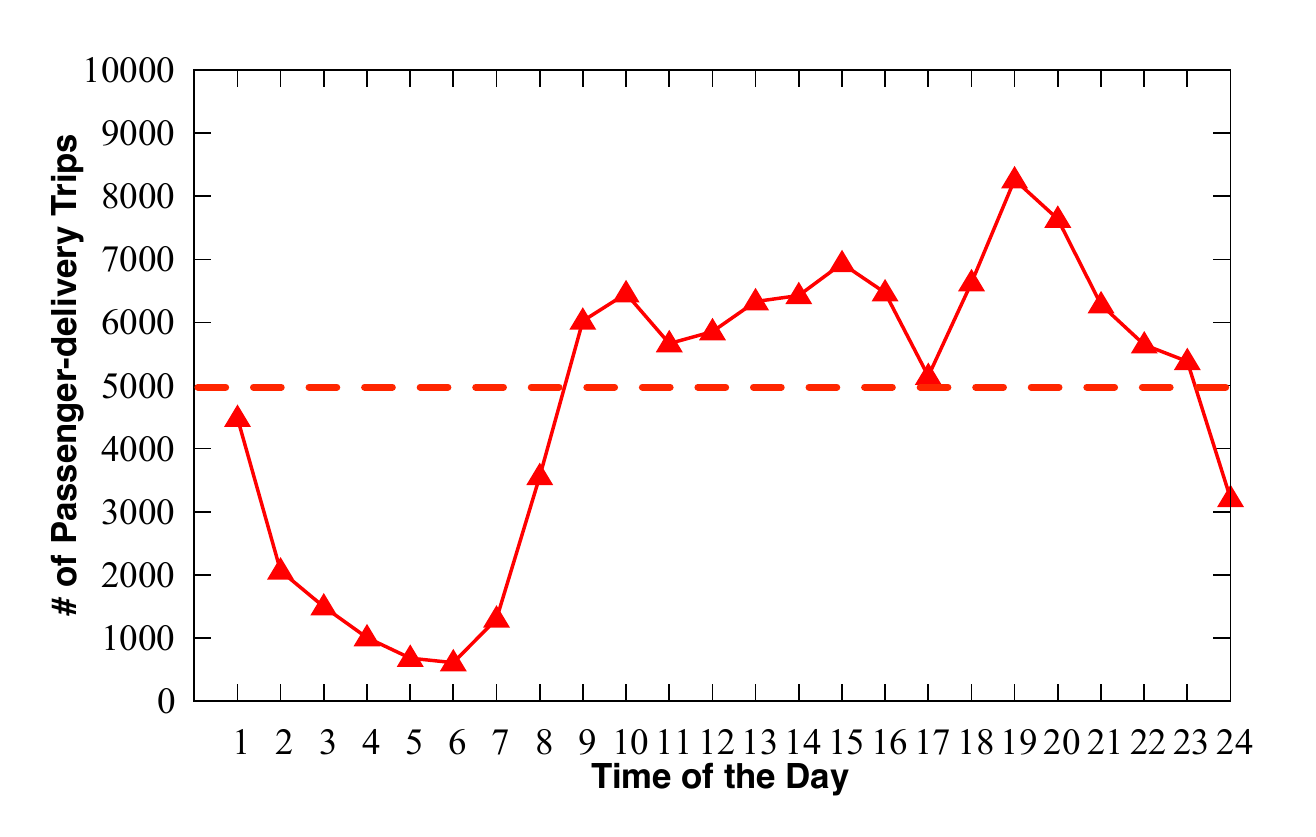}\label{fig:trips}}
    \caption{Evaluation results of success rate under different birth times of the package deliveries.}
\end{figure}

We are also interested at the performance of the success rate at different time of the day. As can be observed in Fig.~\ref{fig:srhours}, the success rate is quite {\em stable and high}  from 8:00 to 18:00, with the value above 95\%, demonstrating the usefulness of the proposed system in practice during the day time. The success rate is extremely low during late-night and early-morning. For instance, the success rate is even lower than 10\% from 2:00 to 5:00. To get an in-depth understanding on the root cause, we further show the number of passenger-delivery trips at different time of the day in Fig.~\ref{fig:trips}.  As expected, the success rate is highly correlated with the number of passenger-delivery trips, i.e., a greater number of passenger-delivery trips implies a higher success rate since more hitchhiking opportunities for package deliveries are provided. An interesting observation is that the success rate at 17:00 is still high though the number of passenger-delivery trips is not large at that time, compared to nearby hours. This is because: on one hand, the number of hitchhiking opportunities for package deliveries should be accumulated during the given deadline (usually bigger than an hour); on the other hand, the number of passenger-delivery trips in the next two hours increases and remains high.  On the contrary, the number of passenger-delivery trips during late-night and early morning is {\em continuously} small. During that time periods, although hitchhiking opportunities are accumulated, it is still {\em insufficient}, resulting in a poor success rate. In this study, the number of package delivery requests is 1,000 for each hour time; the $extraT$ is fixed to 60 minutes.

\begin{figure}
    \centering
    \includegraphics[width=0.884\columnwidth]{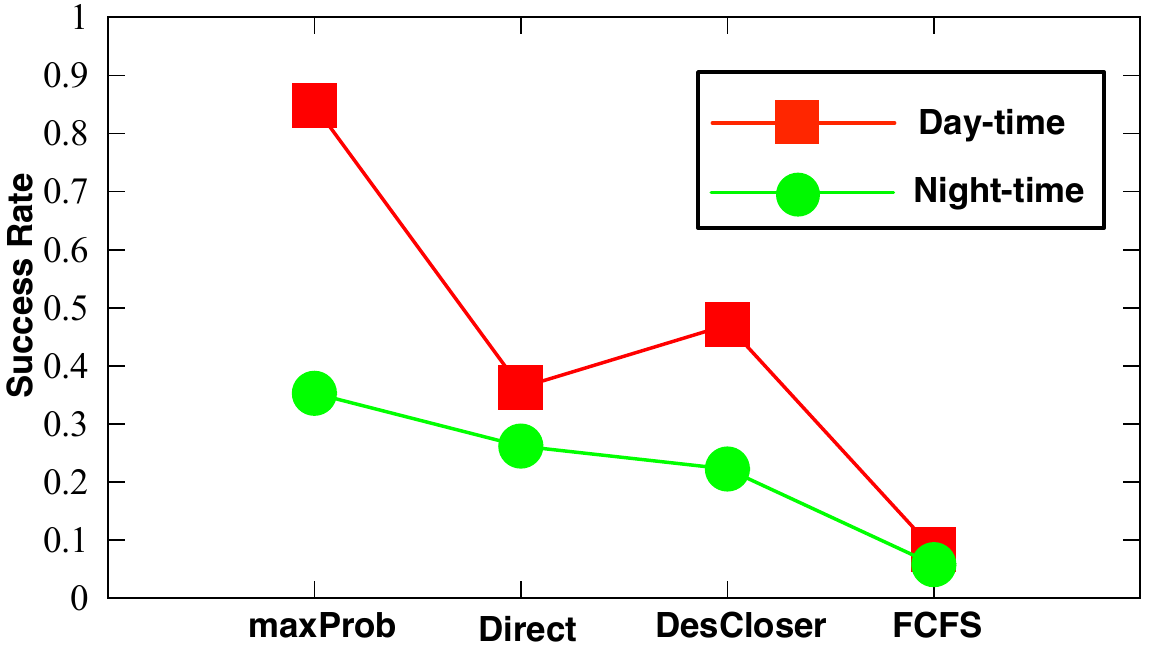}
    \caption{Evaluation results of success rate for four different  algorithms at day time and night time respectively.}
    \label{fig:srAlgorithms}
\end{figure}

We report the result of success rate for four different algorithms  w.r.t  the time of the day. For simplicity, we do not provide the success rate of the four algorithms under every hour of a day, and just split a day into two time slots, i.e, day time from 8:00 to 18:00 and the rest is the  night time, respectively. In the time dimension, as shown in Fig.~\ref{fig:srAlgorithms}, for all four algorithms, the success rate is higher at day time than night time, except for {\bf FCFS} which only achieves the similarly low success rate at both time slots. {\bf Direct} algorithm returns quite close performance at day time and night time. As predicted, the success rate of {\bf FCFS} algorithm is the lowest, i.e., below 10\%.  Compared to the other three baseline algorithms, {\bf  maxProb} achieves the best success rate at both day time and night time. In this experiment, the number of package delivery requests is 10,000, with the birth time uniformly distributed at the day time and nigh time, respectively; the $extraT$ is fixed to 60 minutes.

In real situation,  there are more passengers who prefer to  take taxis at some interchange stations, providing more hitchhiking opportunities for  package deliveries, probably leading to a better success rate. We thus manually categorize the interchange stations into two classes (i.e. {\em popular, unpopular}) in advance, by taking its total number of pick-ups and drop-offs in history  into account. The number of interchange stations labeled as {\em popular} is 19; and the number of interchange stations labeled as {\em unpopular} is 15.  We further identify three categories of package delivery requests, according to the labels of the original and destination  station, as shown in Table~\ref{tab:categories}. Then, we test the performance on success rate for each category. The number of package delivery requests for each category is same, with a value of 10,000. 

\begin{figure}
    \centering
    \includegraphics[width=0.884\columnwidth]{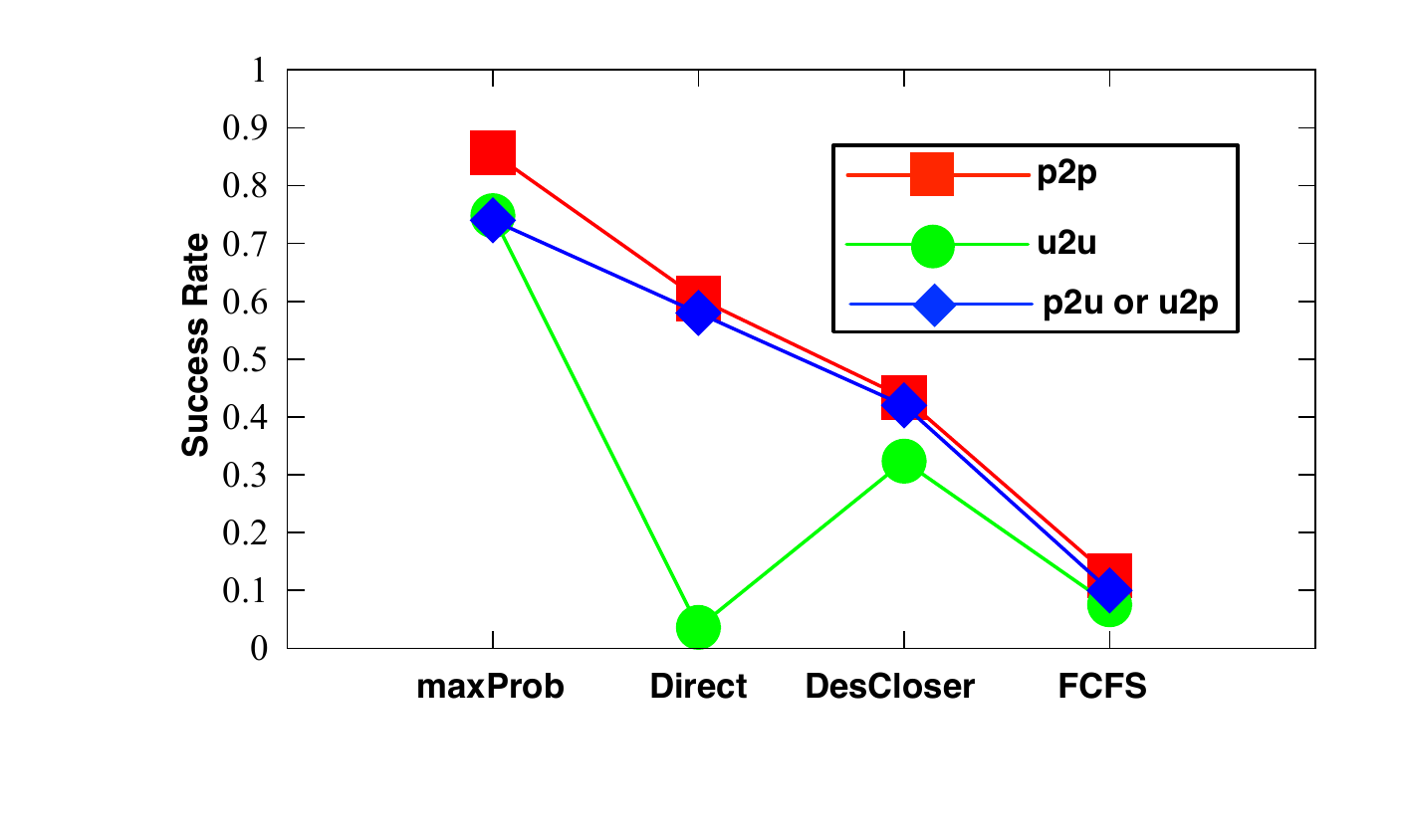}
    \caption{Evaluation results of success rate for four different algorithms of different package categories.}
    \label{fig:srAlgorithmsRegions}
\end{figure}

\begin{table}[h]
\centering
\caption{Three identified categories of package delivery requests.}
\label{tab:categories}
\begin{tabular}{c||c}
\toprule
  &  Description  \\
 \hline\hline
 Category I       & {\em packages start and end at  popular stations}   (p2p)           \\
 Category II        & {\em packages start or end at  popular stations}   (p2u or u2p)           \\
 Category III       & {\em packages start and end at unpopular stations}  (u2u)            \\
\bottomrule                   
\end{tabular}
%\vspace{-0.4cm}
\end{table}

Fig.~\ref{fig:srAlgorithmsRegions} shows the comparison of success rate of the four algorithms under a given  deadline (the $extraT$ is fixed to 60 minutes in this study) for the three  categories. {\bf maxProb} achieves the best performance for three categories, while the performance of {\bf FCFS} is the worst. One exception is   the success rate of {\bf Direct} for the Category III packages. In such case, without any relay at the intermediate stations, {\bf Direct} obtains an extremely low success rate (i.e., under 5\%), which is due to the fact that there is insufficient passenger flow between two unpopular stations. For the same algorithm, the performance is also different for different categories of package delivery requests. The performance for the Category I packages is the best; the performance is the worst for the Category III packages.  Particularly,  {\bf maxProb} ensures that around 90\% of the Category I packages  (75\% for Category II and III) can be arrived-on-time successfully.  While for {\bf FCFS}, only less than 10\% of all Category packages can be delivered by the deadline. Similar to the the previous results, {\bf DesCloser} performs better than {\bf FCFS} and worse than {\bf maxProb} for all three categories.  %The number of package delivery requests  is the same (i.e. 5000) for all three categories.  As shown in Table~\ref{tab:requests},  the average driving distance of the Category I packages is the smallest (i.e. 12.998). In comparison, the average driving distance of the Category II and III is 25.453 and 29.075 kilometers, respectively.  % with our proposed {\bf AdaPlan} algorithm, around 90\% of them around 75\% of Category II packages can reach destinations within 6 hours; around 60\% of Category III packages can arrive at destinations less than 6 hours.

\subsubsection{Results of  Number of Relays}\label{sec:relayResult}
\begin{figure}
    \centering
    \includegraphics[width=0.884\columnwidth]{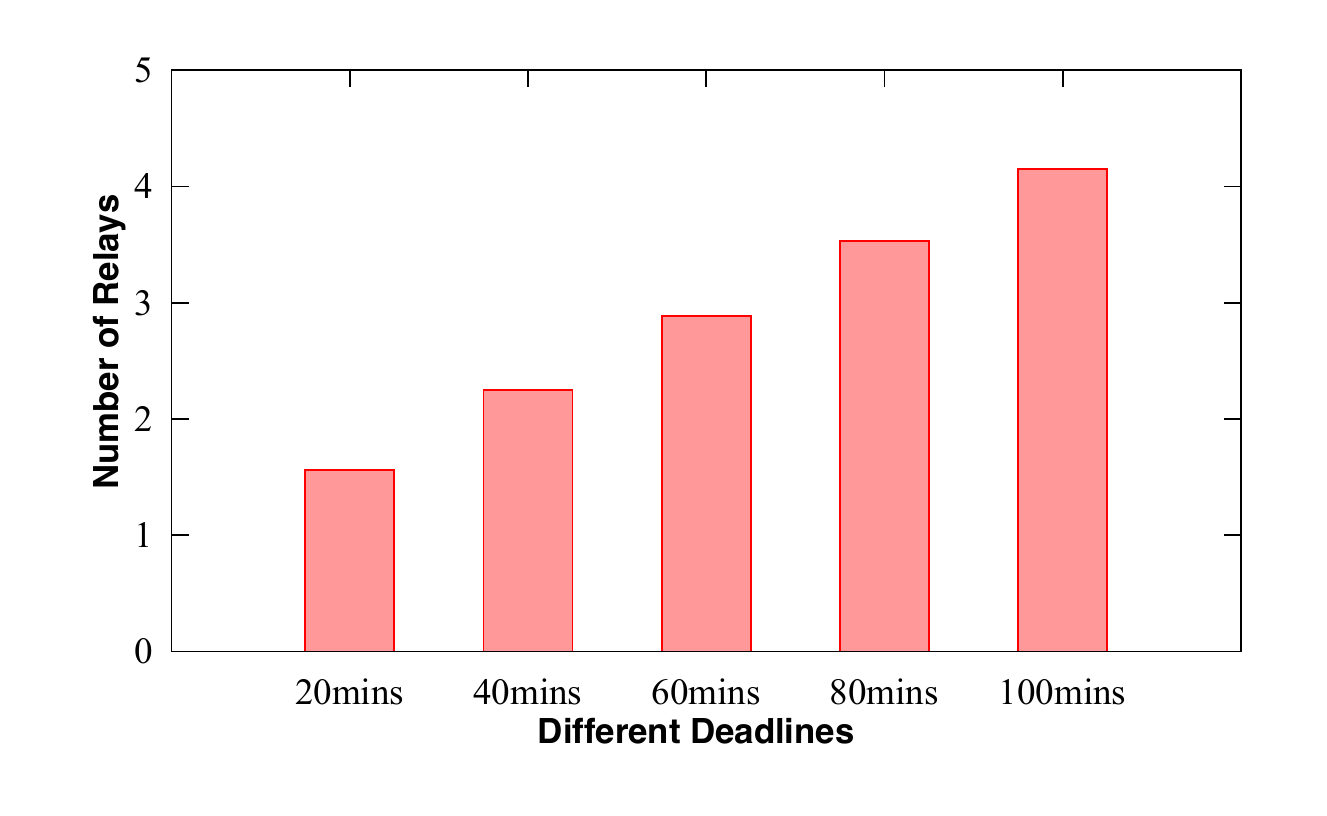}
    \caption{Evaluation results of the number of relays under different deadlines.}
    \label{fig:relaydeadlines}
\end{figure}

We compare the number of relays\footnote{For the experiments on the number of relays, it should be noted that the number of relays is counted and averaged only for the packages that are delivered on time.} of {\bf maxProb} under different given deadlines, with the results shown in Fig.~\ref{fig:relaydeadlines}.  One can observe that the number of relays presents an ascending tendency with the increase of the given deadline. One possible reason is that more {\em ineffective} relays (i.e., the package moves back and forth towards the destination commonly) are resulted since {\bf maxProb} is inclined to hitchhike the coming taxi immediately, as discussed. 

\begin{figure}
    \centering
    \includegraphics[width=0.884\columnwidth]{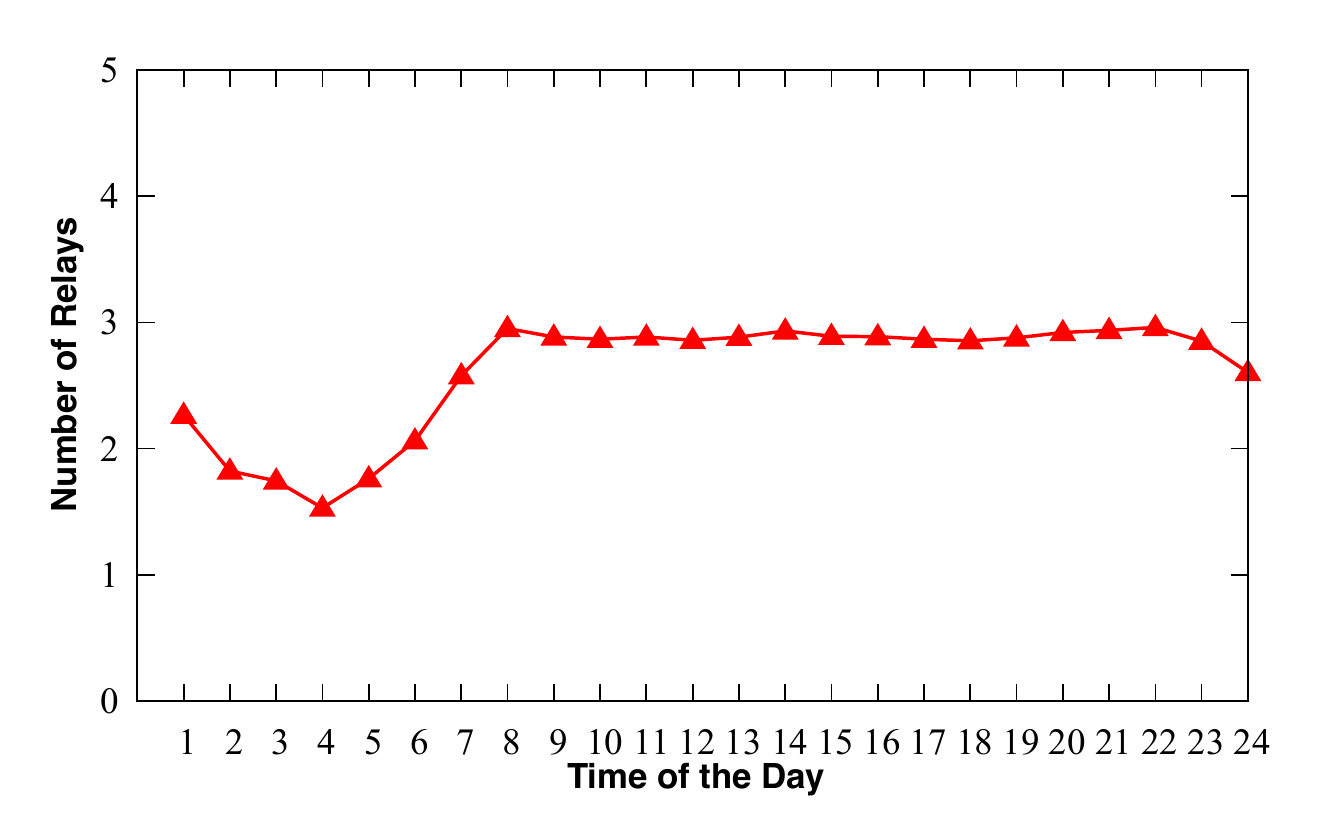}
    \caption{Evaluation results of the number of relays under different time of the day.}
    \label{fig:relaytimes}
\end{figure}

We also show the results on how the number of relays obtained  by {\bf maxProb} varies under  different time of day in Fig.~\ref{fig:relaytimes}. Overall, the number of relays is more or less unchanged (i.e., with a value of around 3) during the whole day, except for the early morning when the number of relays is relatively smaller, probably caused by the extremely little passenger flow during that time. 

\begin{figure}
    \centering
    \includegraphics[width=0.884\columnwidth]{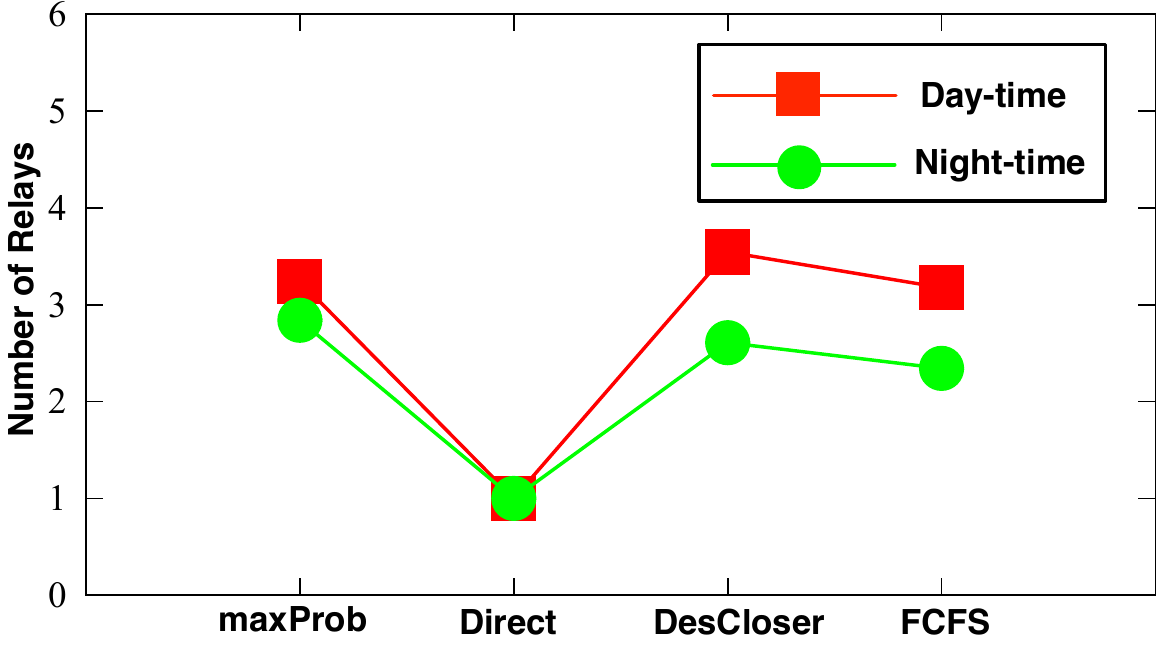}
    \caption{Evaluation results of the number of relays for four algorithms at day time and night time respectively.}
    \label{fig:relayAlgorithmsTime}
\end{figure}

We report the results of the number of relays for four algorithms w.r.t the day time and nigh time respectively, as shown in Fig.~\ref{fig:relayAlgorithmsTime}. As can be predicted, the number of relays for {\bf Direct} shall equal one. For the other three algorithms,  slightly more relays are generally required at the day time than the night time. Moreover, a little surprisedly, the number of relays resulted from {\bf DesCloser} and {\bf FCFS} are quite close to that obtained by {\bf maxProb}, implying that the number of relays  is somehow independent of the adopted algorithms for a successful package delivery. We will investigate deeper about the potential causes qualitatively and quantitatively such as the geographical and temporal distributions of the successful package deliveries in the future work. %To avoid confusion, we also calculate the number of relays for the packages that cannot be delivered on time for {\bf DesCloser} and {\bf FCFS} respectively.  

\begin{figure}
    \centering
    \includegraphics[width=0.884\columnwidth]{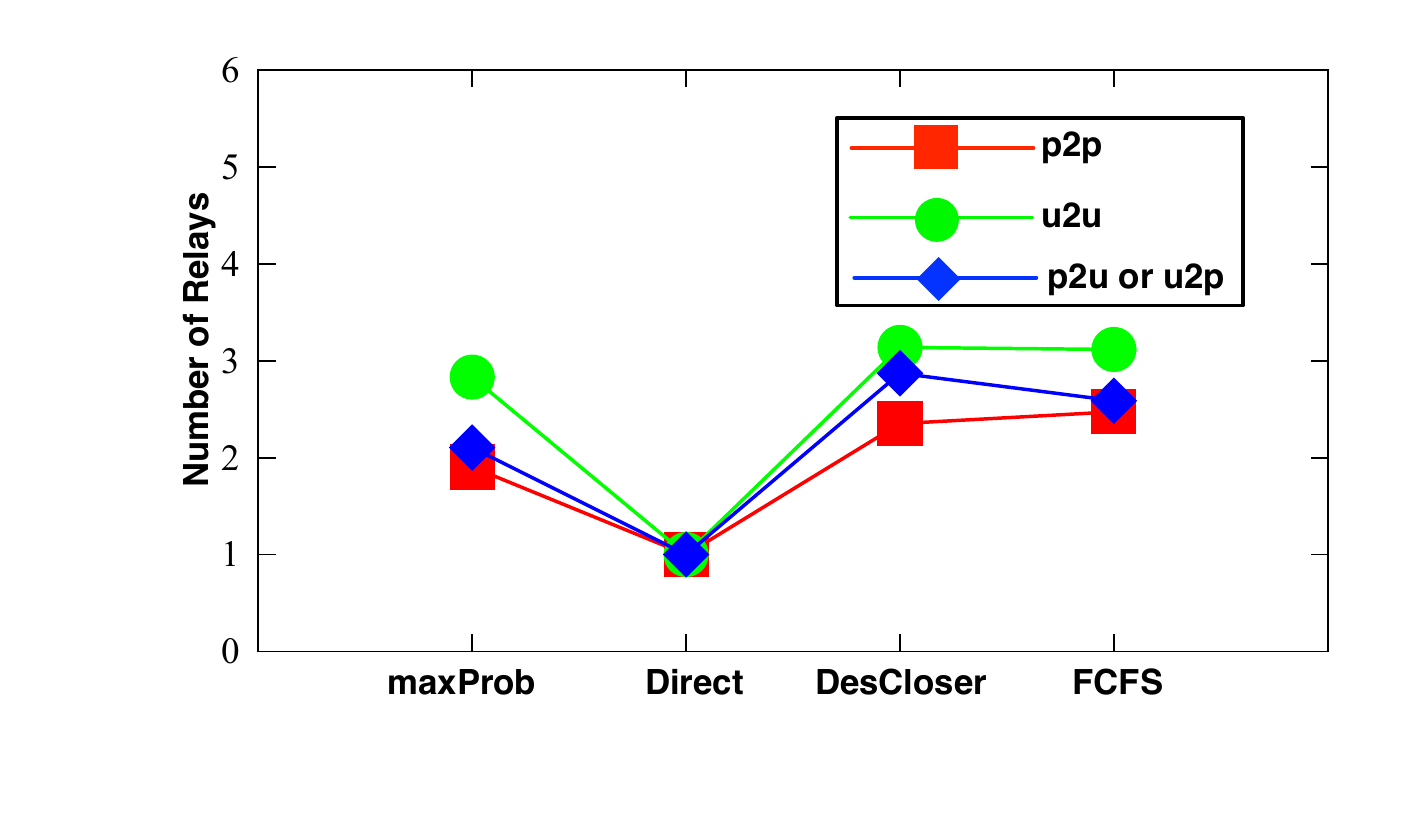}
    \caption{Evaluation results of the number of relays for four algorithms for different package categories.}
    \label{fig:relayAlgorithmsRegion}
\end{figure}

We further report the results of the number of relays for four algorithms w.r.t the package categories, as shown in Fig.~\ref{fig:relayAlgorithmsRegion}. Similarly, the number of relays for {\bf Direct} is one, regardless of the package categories. Compared to the other two algorithms (i.e., {\bf DesCloser} and {\bf FCFS}), {\bf maxProb} requires slightly fewer relays for all three package categories. Similar to the results of the success rate w.r.t the package categories, for all algorithms except for {\bf Direct},  the performance in terms of the number of relays is the best for the Category I packages; the performance for the Category III packages is the worst.   

\subsubsection{Results of  Package Throughput}\label{sec:throughput}
\begin{figure}
    \centering
    \includegraphics[width=0.994\columnwidth]{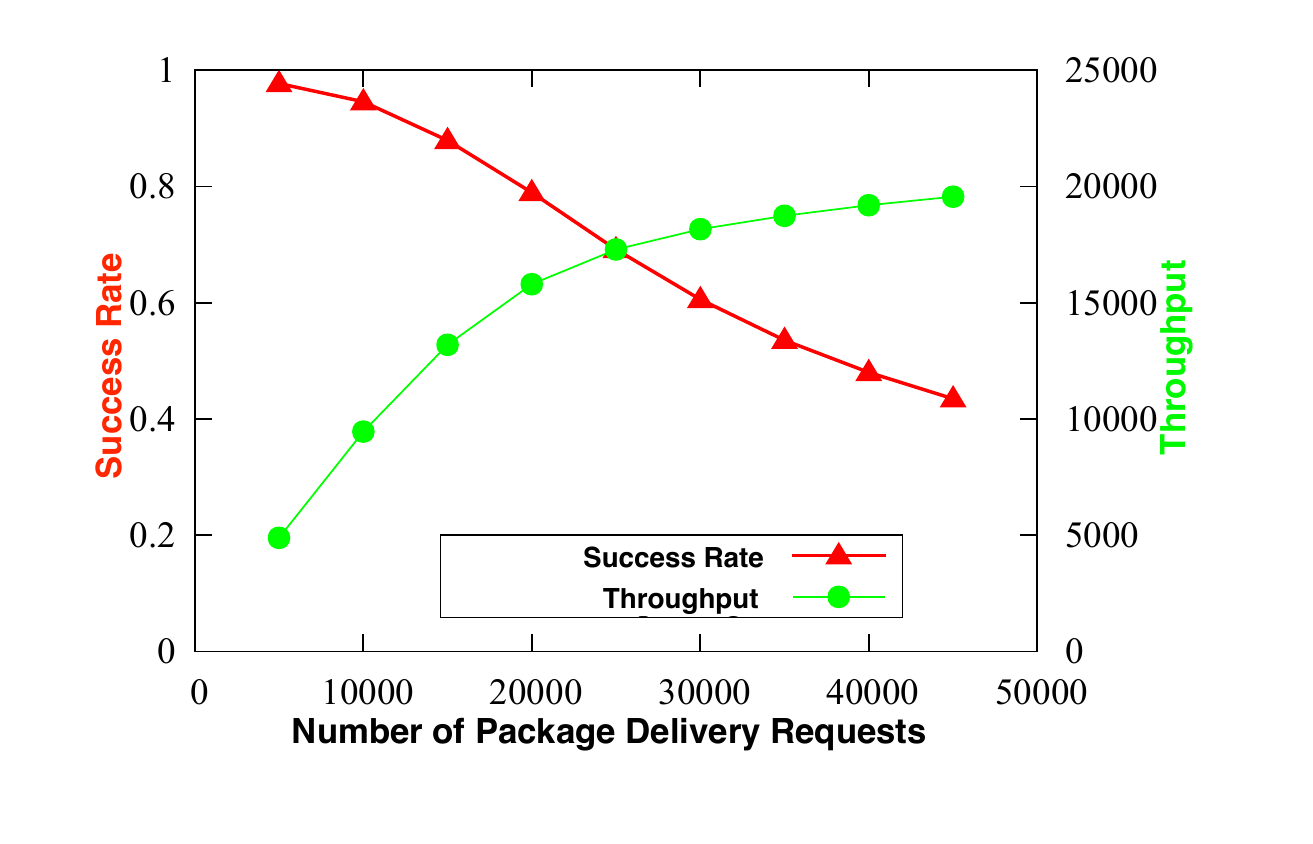}
    \caption{Evaluation results of the number of relays for four algorithms for different package categories.}
    \label{fig:throughput}
\end{figure}

It is necessary to evaluate the package throughput of the system, since it is a primary consideration when applied to real-life scenarios. Fig.~\ref{fig:throughput} shows the results on the number of packages that the proposed system can transport successfully w.r.t the number of total generated package delivery requests per day, together with the success rate. More specifically, package throughput increases gradually with the number of generated package delivery requests before approaching a stable value. On the contrary, the success rate declines almost linearly with the number of generated package delivery requests.  As can be seen, the maximum package throughput is around 20,000 per day, however, the corresponding success rate is quite low (i.e., around 40\%) and might be not applicable in real life.  To make the proposed system practical in real situations, the package throughput should be around 9,500 per day while maintaining a relatively promising success rate.  

\begin{figure}
    \centering
    \includegraphics[width=0.994\columnwidth]{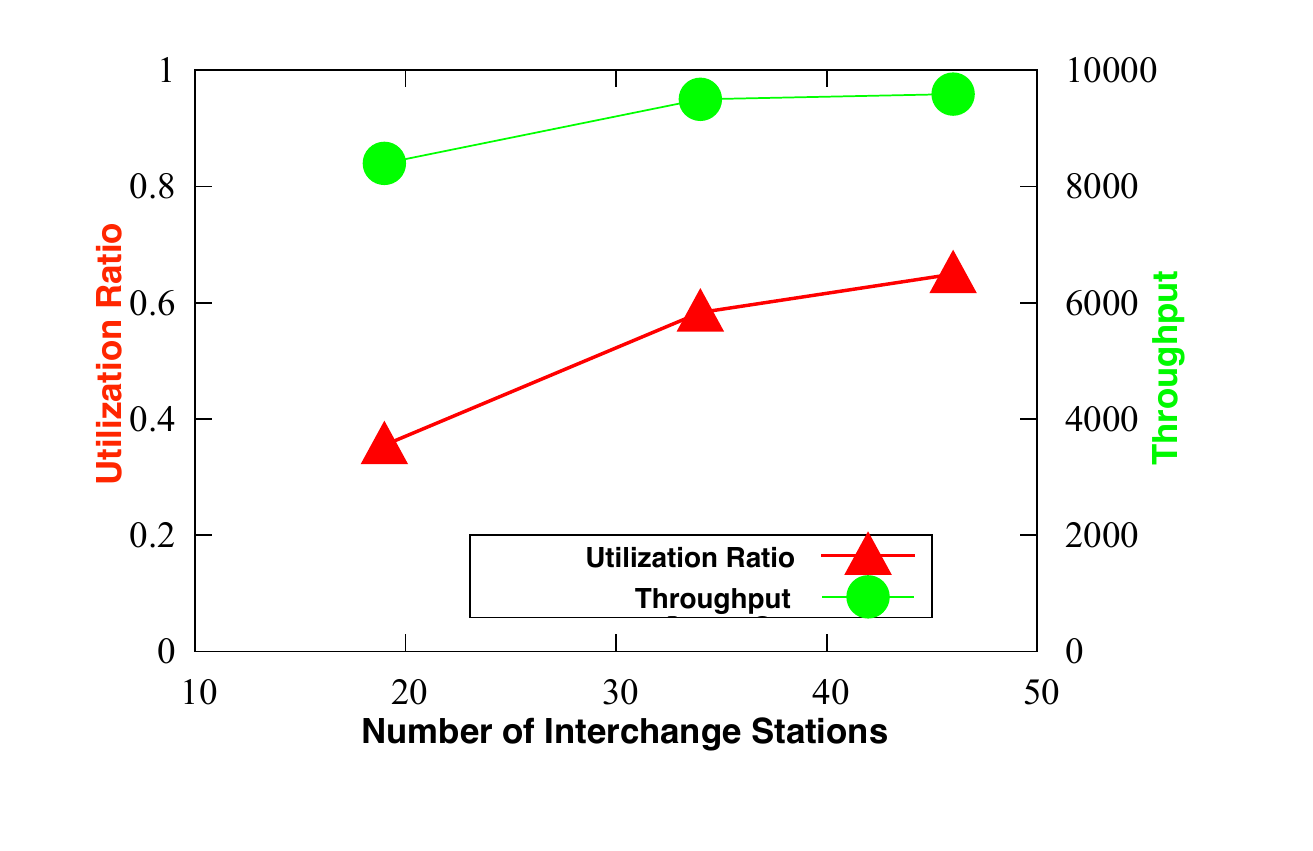}
    \caption{Evaluation results of the number of relays for four algorithms for different package categories.}
    \label{fig:density}
\end{figure}

We further examine the package throughput under different density (number) of interchange stations, as shown in Fig.~\ref{fig:density}. As expected, the more interchange stations our system has, the higher throughput it can achieve (the number of package delivery requests for this study is 10,000). We argue that the {\em root cause is the improvement on utilization ratio of taxi trips for package deliveries}, which is  defined as the ratio between the number of taxi trips involved in package deliveries and the total number of taxi trips. We thus plot the {\em utilization ratio}  under different density of interchange stations in Fig.~\ref{fig:density} as well.  As evidenced, an increasing utilization ratio (35.5\%, 58.3\% and 64.9\% respectively) is achieved as the interchange stations in the city becomes denser (18, 34 and 46 respectively). Moreover, there is still a considerable room to increase the package throughput further if placing more interchange stations to better utilize the taxi trips. We are aware that the choice of interchange stations (different locations but with the same number) may affect the throughput result, but the overall trend shall be the same.  %However, quite limited improvement on throughput can be achieved when the number of interchange station increases from 34 to 46.  is the number of hitchhiking rides provided by the taxis than can be utilized in the city.

\subsection{Open Research Issues}
In this section, we discuss some open issues  which are not resolved in this work but will be  addressed in the future.

{\color{black}
{\bf Package Delivery Request.}  Modeling package delivery request is a challenging task, requiring additional data sources, such as demographics, socio-economics, land use, and on-line purchasing activities. How to accurately model the city-wide package flow distributions can be a separate research problem itself, and there has not yet been any reliable model as far as we know~\cite{gonzalez2014emission}. In the scope of this paper, we focus on discovering the near-optimal package delivery paths with high efficiency, given any package delivery request. Hence, we simply generate the package birth time, origins and destinations randomly.  In the  future, we plan to incorporate other  more real-world data sources to model the city-wide package delivery requests~\cite{tan14express}, and integrate them into our framework.%As expected, online purchasing activity spikes at lunchtime and towards 5pm in the UK. However, activity hits its peak later into the evening, showing that consumers are spending even when high-street stores are closed.

{\bf  Transport Network Optimization.} At the current research of crowd logistics, most studies focus on developing advanced package routing algorithms. In  contrast, less attention has been paid to the package transport network optimization (i.e., the number/locations of interchange stations). In the future, we plan to address the two issues simultaneously.  

{\bf Multi-objective Optimization.} The on-time performance is one of the important optimization objectives of crowdsourced logistics. From the side of  logistics service providers, other objectives are also equally important, such as the money paid to the taxi drivers. The more taxi drivers are recruited, the more rewards are necessary. Thus, to minimize the number of package relays is worth to be considered. From the side of taxi drivers, the minimization of detour driving distance is of great importance. In the future, we plan to formulate the problem of crowd delivery via hitchhiking taxi rides  as the multi-objective optimization one. 

{\bf  Practical Issues.} There are still many practical issues to be addressed before truly realizing the system. The capacity issue of the interchange station  is one example. On one hand, interchange stations  frequently visited by passengers are more likely to be recruited for relaying packages,  and thus require a larger capacity in general. On the other hand, the spatial and temporal traffic patterns of package deliveries also have a critical impact on the capacity issue~\cite{crainic2014logistics}.  Other practical issues such as the incentive mechanism for taxi drivers, the package pricing methods, the cooperation cost of the interchange station, the size of the package, the standard for container of the package  also need further investigation. 

}

\section{Conclusion and Future Work}\label{sec:conclusion}
In this paper, we present a novel framework called CrowdExpress for package delivery path planning.  The framework proposes to exploit {\em hitchhiking rides} provided by occupied taxis  to transport packages in time without degrading the quality of passenger services.  More specifically, we first built the package transport network by mining the taxi GPS trajectory data offline. Then we proposed a two-phase approach for package delivery path planning with a novel and comprehensive  process. Using  real-world datasets which include  road network data, and a large-scale taxi trajectory data generated by over 19,000 taxis in a month in NYC, US, we compared our proposed method with three baseline algorithms, and showed that our method is more efficient and effective. 

In the future, we plan to broaden and deepen this work in several directions. First, we plan to correlate the package deadline to the package pricing. Currently, we set a general and relative deadline for all package deliveries and handle them equally. In the near future, we plan to set different priorities for packages according to users' expected arriving time. For instance, users are charged higher if they want their packages to be arrived earlier,  and new package routing algorithms should be developed accordingly.  Second, we intend to take actions to improve the system throughput and coverage. Such as grouping packages with close destinations and optimizing the interchange stations (number and location).  Finally, we plan to implement and test our system with real users in actual settings, collecting feedback on how to further improve the service.

% re-examine some assumptions made. For instance, in real life, the assumption that taxi drivers are willing to accept the package delivery task may not always hold. We thus need to consider more realistic assumptions.  Second, we plan to exploit more hitchhiking opportunities provided by other transportation modes, such as private cars, bicycles and public buses. Third, we would like to deploy our system on mobile devices, enabling a series of pervasive services for taxi drivers and on-line shoppers. Finally, we plan to test our system with real users in actual settings, collecting feedback on how to further improve the service.

\section*{Acknowledgements}
The work was  supported by the National Key Research and Development Project of China (No. 2017YFB1002000), the  National Science Foundation of China (No. 61602067 and No. 61872050),  Chongqing Basic and Frontier Research Program (No. cstc2018jcyjAX0551) and the Fundamental Research Funds for the Central Universities (No. 2018cdqyjsj0024). Chao Chen is the corresponding author.
%\section*{Reference}
\bibliographystyle{abbrv}
\bibliography{taxiFed.bib}
%\newpage

\begin{IEEEbiography}[{\includegraphics[width=1in,height=1.25in,clip,keepaspectratio]{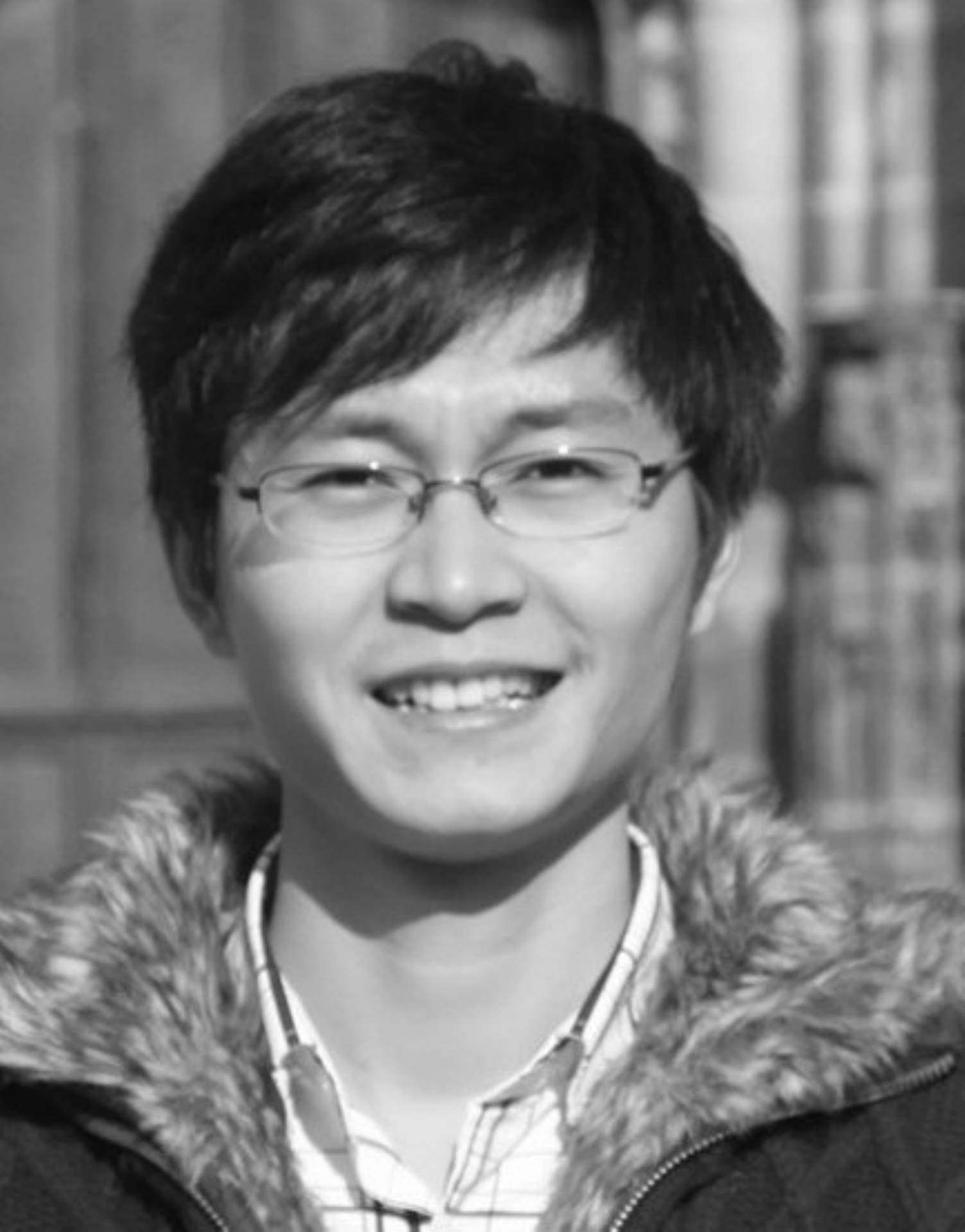}}] {Chao Chen} is an associate professor at College of Computer Science,  Chongqing University, Chongqing, China. He obtained his Ph.D. degree from Pierre and Marie Curie University and  Institut Mines-TELECOM/TELECOM SudParis, France in 2014. He received the B.Sc. and M.Sc. degrees in control science and control engineering from Northwestern Polytechnical University, Xi’an, China, in 2007 and 2010, respectively. Dr. Chen  was the recipient of  the Best Paper Runner-Up Award at MobiQuitous 2011.

In 2009, he worked as a Research Assistant with Hong Kong Polytechnic University, Kowloon, Hong Kong. His research interests include pervasive computing, social network analysis, urban logistics, data mining from large-scale taxi data, and big data analytics for smart cities. 
\end{IEEEbiography}

\begin{IEEEbiography}[{\includegraphics[width=1in,height=1.25in,clip,keepaspectratio]{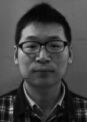}}]{Sen Yang}is currently a master student at College of Computer Science,  Chongqing University, Chongqing, China.  His research interests include scenic travel route planning, taxi GPS trajectory data mining, and last-mile delivery. 
\end{IEEEbiography}

\begin{IEEEbiography}[{\includegraphics[width=1in,height=1.25in,clip,keepaspectratio]{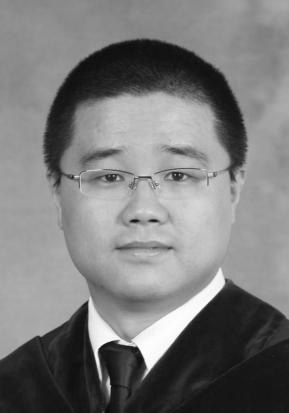}}] {Weichen Liu} (S’07-M’11) is currently a professor at College of Computer Science, Chongqing University, China. Before that, he worked as a postdoctoral research fellow with teaching duties in Nanyang Technological University, Singapore. He received the PhD degree from the Hong Kong University of Science and Technology, Hong Kong, and the MEng and BEng degrees from Harbin Institute of Technology, China. 

Dr. Liu served as the chairs, technical program committee members, editors and reviewers for over 20 premier international conferences and journals. He
authored and co-authored more than 70 research papers in peer-reviewed journals, conferences and books, and received the best paper candidate awards from ASP-DAC 2016, CASES 2015, CODES+ISSS 2009, the best poster award from AMD-TFE 2010, and the most popular poster award from ASP-DAC 2017. His research interests include embedded and real-time systems and network-on-chip. 
\end{IEEEbiography}

\begin{IEEEbiography}[{\includegraphics[width=1in,height=1.25in,clip,keepaspectratio]{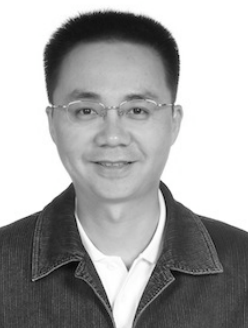}}]{Yasha Wang}  received his Ph.D. degree in Northeastern University, Shenyang, China, in 2003. He is a professor and associate director of National Research \& Engineering Center of Software Engineering in Peking University, China. His research interest includes urban data analytics, ubiquitous computing, software reuse, and online software development environment. He has published more than 50 papers in prestigious conferences and journals, such as ICWS, UbiComp, ICSP and etc. As a technical leader and manager, he has accomplished several key national projects on software engineering and smart cities. Cooperating with major smart-city solution providing companies, his research work has been adopted in more than 20 cities in China. 
\end{IEEEbiography}

\begin{IEEEbiography}[{\includegraphics[width=1in,height=1.25in,clip,keepaspectratio]{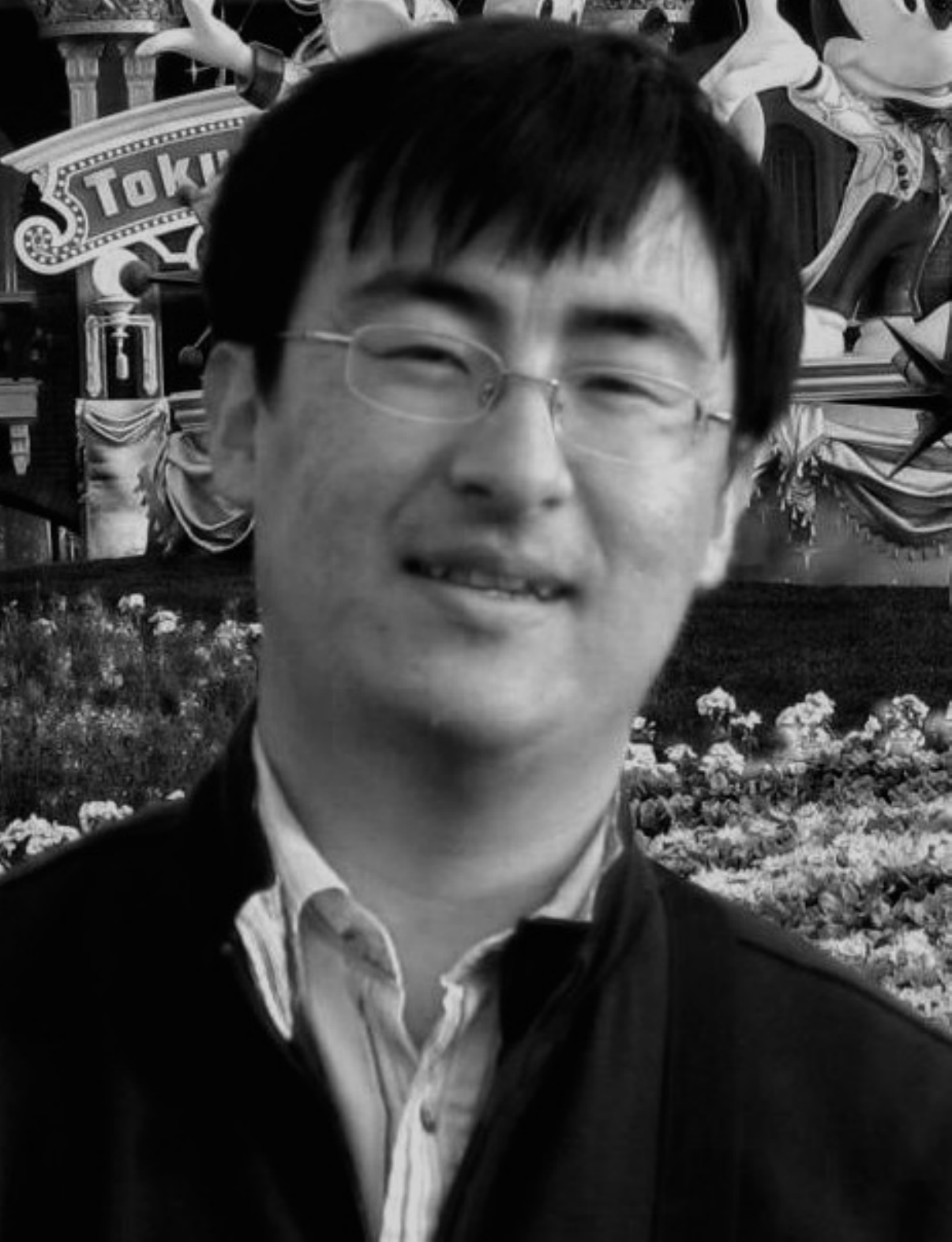}}]{Bin Guo}  is a professor from Northwestern Polytechnical University, China. He received his Ph.D. degree in computer science from Keio University, Japan in 2009 and then was a post-doc researcher at Institut Mines-TELECOM/TELECOM SudParis in France. His research interests include ubiquitous computing, mobile crowd sensing, and HCI.
\end{IEEEbiography}

\begin{IEEEbiography}[{\includegraphics[width=1in,height=1.25in,clip,keepaspectratio]{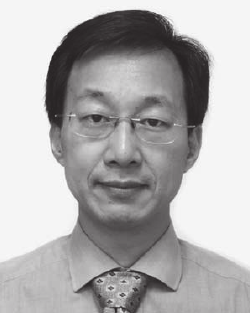}}]{Daqing Zhang} is a professor at  Institute of Software, School of Electronics Engineering and Computer Science, Peking University, China. He obtained his Ph.D from University of Rome ``La Sapienza'' and the University of L’Aquila, Italy in 1996. His research interests include large-scale data mining, urban computing, context-aware computing, and ambient assistive living.  He has published more than 180 referred journal and conference papers, all his research has been motivated by practical applications in digital cities, mobile social networks and elderly care. 

Dr. Zhang is the Associate Editor for four journals including {\em ACM Transactions on Intelligent Systems and Technology}. He has been a frequent Invited Speaker in various international events on ubiquitous computing. He is the winner of the Ten Years CoMoRea Impact Paper Award at IEEE PerCom 2013, the Best Paper Award at IEEE UIC 2015/2012 and the Best Paper Runner Up Award at Mobiquitous 2011. He is a member of the China Thousand-Talent Program.
\end{IEEEbiography}

\appendices
\section{Proof of Theorem 1}\label{sec:appProof}
\begin{proof}
The  theorem can be proven by induction, detailed as follows:

{\bf Base Case:} When $n = 1$ (in which $n$ is length of the given path which is quantified by the number of interchange stations contained by the path minus one), it is obviously that we have $P_{ij}(t\le t_1|Path^{n=1})\le P_{ij}(t\le t_2|Path^{n=1}),~if~t_1<t_2$, according to {\em Definition}~\ref{def:probability}.

{\bf Induction Step:} Let $k\in N$ be given and suppose Eq.~\ref{eq:theorem} is true for $n=k$, that is, we have:
$$P(t_2|Path^{n=k})\geq P(t_1|Path^{n=k}),~if~t_1<t_2$$ 
Then,
$$P(t_1|Path^{n=k+1})= \int_{0}^{t_1}p(t)P(t_1-t|Path^{n=k})dt$$
where $p(t)$ is the probability density function on the edge from the origin to the first stop. 
\begin{equation}
\begin{aligned}
P(t_2|Path^{n=k+1}) = \int_{0}^{t_2}p(t)P(t_2-t|Path^{n=k})dt\\
= \int_{0}^{t_1}p(t)P(t_2-t|Path^{n=k})dt+\\ \int_{t_1}^{t_2}p(t)P(t_2-t|Path^{n=k})dt\\
\geq \int_{0}^{t_1}p(t)P(t_1-t|Path^{n=k})dt +\\ \int_{t_1}^{t_2}p(t)P(t_2-t|Path^{n=k})dt\\
= P(t_1|Path^{n=k+1})+ \int_{t_1}^{t_2}p(t)P(t_2-t|Path^{n=k})dt\\
\geq  P(t_1|Path^{n=k+1}) 
\end{aligned}   \nonumber
\end{equation}
    
{\bf Conclusion:} By the principle of induction, Eq.~\ref{eq:theorem} is true for all $n\in N$. 
\end{proof}
\end{document}